\documentstyle[preprint,aps]{revtex}
\begin{document}

\title{
Quantum Electrodynamics of Internal Source X-Ray Holographies:
Bremsstrahlung, Fluorescence, and Multiple Energy X-Ray Holography}

\author{Gerald A. Miller and Larry B. Sorensen}
\address{Department of Physics, University of Washington, Seattle,
Washington 98195}
\maketitle
\begin{abstract}

Quantum electrodynamics (qed) is used to derive the differential
cross sections measured in the three new experimental internal
source ensemble x-ray holographies: bremsstrahlung (BXH),
fluorescence (XFH), and  multiple-energy (MEXH) x-ray holography.
The polarization dependence of the BXH cross section is also
obtained.  For BXH, we study  analytically and numerically the
possible effects of the virtual photons and electrons which enter
qed calculations in summing over the intermediate states. For the
low photon and electron energies used in the current experiments,
we show that the virtual intermediate states produce only very
small effects. This is because the uncertainty principle limits
the distance that the virtual particles can propagate to be 
much shorter than the separation between the regions of
high electron density in the adjacent atoms. We also find that
using  the  asymptotic form of the scattering wave function
causes about a 5-10\% error  for near forward scattering.

\end{abstract}
\vspace {1cm}

\hfill {DOE/ER/41014-4-N97}

\hfill {DOE/ER/40561-300-INT96-00-154}

\newpage
\section{Introduction}

Fifty years ago, 
Gabor proposed electron holography as a method to
improve the resolution of electron microscopes so that atoms could be
directly imaged \cite{gabor}.  Gabor's idea was to focus the electron beam 
to a very small region of space just outside the sample to produce a nearly
point source of radiation, and to record the interference pattern between the
spherical reference wave from this point source and the spherical object waves 
produced when the electrons scattered from the atoms in the sample.  This
photographically recorded interference pattern would then be used as the 
diffraction grating in an optical reconstruction system.  Although
Gabor's dream to directly image atoms using 
holographic electron microscopy has never been realized (because the
quality of the best electron lenses is only about as good as that of a raindrop 
for visible 
light \cite{gabor2}), Gabor's suggestion produced the optical holography
revolution with the 
advent of lasers to provide the necessary coherent monochromatic
external reference waves.

Ten years ago, Sz\"oke pointed out that the necessary coherent spherical 
reference wave
could  also be created by generating 
the electron reference wave inside the sample
\cite{S86}.  In this case, the spatial coherence comes from the small spatial
extent of the internal electron source.
Sz\"oke's internal source electron holography suggestion generated a flurry of 
activity \cite{so_many},
and, in the past five years, 
Gabor's dream of directly imaging atoms with electrons has 
been realized using photoelectrons \cite{photoe},
Auger electrons \cite{auger}, 
diffusely scattered low energy electrons \cite{LEED},
and diffusely scattered Kikuchi electrons \cite{Kik}.
However, because electrons 
interact very strongly with atoms, the scattered object
waves are not very good spherical waves (there is a strong angular variation of
the magnitude and the phase of the  electron--atom scattering amplitude), 
and multiple scattering produces
``electron focussing" effects along the lines of atoms in the sample
which are important.
Consequently, these new electron holographies produce pictures of where the 
atoms are,
but they do not accurately reconstruct the atomic positions. 
Sz\"oke also proposed internal source x-ray holography using 
fluorescence x-rays.
Because x-rays 
interact weakly with atoms, internal source x-ray holograms should
produce much more accurate atomic resolution images than the 
internal source electron holographies \cite{early_tegze,charholo}.  
Unfortunately, the price for this is that the modulation of the
intensity in x-ray holograms ($1-5 \times 10^{-3}$) is 
about 100 times weaker than the 
modulation 
in electron holograms ($1-5 \times 10^{-1}$).

Earlier this year, the first atomic resolution x-ray holograms were
produced using  x-ray fluorescence holography (XFH) \cite{XFH}
and multiple energy x-ray holography (MEXH) \cite{MEXH}.  
About three years ago, 
stimulated by 
the advantages of multiple energy electron holography \cite{multiple}
and the promise of XFH \cite{early_tegze},
we started developing a new kind of 
internal source x-ray holography which
uses bremsstrahlung photons created inside the sample
\cite{silvia}.  
The  primary motivation for this paper is to provide the
theoretical foundation for experimental bremsstrahlung
 x-ray holography (BXH) starting from quantum
electrodynamics. The bulk of this paper is devoted to BXH, but we also 
show how the same quantum electrodynamic foundation applies to XFH and MEXH.

Bremsstrahlung x-ray holography is very attractive for three reasons:
(1) Bremsstrahlung allows hard x-rays to be produced from low Z atoms.
High quality holograms require the wavelength to be much smaller than the
spacing between the atoms.  The characteristic fluorescence energies of
many interesting and important low Z elements are too low to provide good
images using XFH.  
(2) Bremsstrahlung 
produces x-rays with a wide spread in their energy and allows
multiple energy holograms to be recorded simultaneously by energy analyzing the
bremsstrahlung photons.
To accurately reconstruct a three-dimensional object in real space, we need
information over a three-dimensional volume in reciprocal space.  To overcome
the problems in the internal source single-energy electron holographies, several
multiple energy                                              
 methods have been developed \cite{multiple}.  These multiple-energy
electron methods eliminate the twin images, greatly reduce the effects due to
the strong angular variation of the magnitude and phase of the electron--atom 
scattering amplitude, and reduce the noise in the reconstructions.
The MEXH method was developed in analogy to the multiple energy electron methods
to provide higher quality holograms than the single-energy XFH method.  
(3) The bremsstrahlung production cross section is extremely high.  
A conventional
400 watt x-ray source produces about $10^{13}$ short wavelength bremsstrahlung
photons per second into $4\pi$ steradians \cite{silvia}.
If all of these photons 
could be collected and energy analyzed, a very high quality 
BXH could be generated with a tabletop apparatus in a few hours.

The implementation of bremsstrahlung x-ray holography  raises a number of
interesting theoretical questions. We first recall that 
bremsstrahlung photons 
can have any energy from
nearly zero to the energy of the incident electron,
and the spectral intensity diverges at low energies: 
$I(\omega)  d\omega \sim \omega^{-1} \; d\omega$.
The bremsstrahlung photons, which
scatter in the target crystal to produce  the object waves, are intermediate
or virtual particles. The momentum of 
each intermediate bremsstrahlung photon can take on any value; 
we must integrate
over all virtual momenta in computing  the scattering amplitude.
Of course, this must occur
for any intermediate particle that produces an
object wave, but
the possible problems are potentially more serious here for the 
virtual bremsstrahlung photons because of the broad nature of the
bremsstrahlung  spectrum.

More generally we should ask:
Are quantum mechanical effects ever important in internal source x-ray
holography?  Or does  the simple wave picture always work?  If a quantum
mechanical approach is needed,
what is the correct quantum mechanical description
of internal source x-ray holography?
When can multi-path photon interference be treated by the scalar wave equation
approximation to Maxwell's equations instead of the full theory of quantum
electrodynamics? 
To answer these questions,
we develop a quantum electrodynamic
treatment 
of the three
internal source x-ray holographies, BXH, XFH and MEXH,
and  compare it in detail with
the simple wave picture.

It is  useful to provide this connection between the fundamental 
theory (QED) and these new holographies.
Almost all of the work in this field has been based on the simple
wave picture.
For example,  Barton's original holographic inversion procedure
\cite{B88} for electrons is based on the Helmholtz-Kirchoff
inversion procedure for classical scalar waves. However,
in the non-relativistic limit, Rous and Rubin\cite{RR}
have recently shown how the Lippmann-Schwinger equation can be used to provide
solutions to the Schrodinger equation
which correctly describe the physics of the
single-energy electron internal source holographies.

For BXH in particular, classical electrodynamics will not produce
the correct answer at high energies because the intermediate photons and
electrons are virtual: the square of their four momenta may not be equal to 
the square of their 
rest masses, ${\rm{p}}^2 \ne {\rm{m}}^2$.
The physics can be divided into on-the-mass-shell amplitudes (called the
``on-shell" or ``real" processes) when ${\rm{p}}^2 = {\rm{m}}^2$, 
and off-the-mass-shell
 amplitudes (called ``off-shell" or ``virtual" processes)  
when ${\rm{p}}^2 \ne {\rm{m}}^2$.  
We show by 
explicit calculation that the virtual photons and virtual electrons do not 
propagate over the entire distance between the regions of high electron density
in two adjacent atoms, and 
consequently classical electrodynamics predicts the right 
behavior. The reason for this comes from the uncertainty 
principle. The amount of off-shellness, or virtuality,
of photons of energy $k_0$ and momentum 
$\vec k$  is measured by the quantity $k_0^2 
-\vec k^2\equiv Q^2$.  If $k_0>|\vec k|$, the virtual photon is
not massless and its range is $Q^{-1}$ 
which is 
much smaller than the interatomic spacing $a$. If $k_0<|\vec k|$, $Q^2$
is negative and $\exp(i |Q| r)$ oscillates rapidly for $r\sim a$ and 
any important contributions are cancelled.
 
Thus our main result is that for real atoms in real solids, excited to emit 
bremsstrahlung or fluorescence x-ray radiation, classical electrodynamics 
works very well because there is no significant overlap between the 
regions of high electron density in the adjacent atoms.
However, because the intermediate state photons and electrons in the
internal source x-ray holographies  
are virtual, it is important to use quantum electrodynamics 
to derive the equations necessary to analyze the holograms.
We provide that derivation for BXH, XFH and MEXH. 

The separated atom approximation, that we use to show that the
full quantum electrodynamic treatment reduces to the classical electrodynamic
expressions for real 
atoms in real solids, is formally analogous to the separated
scatterer approximation used 
in analyzing high energy hadron-nuclear
scattering experiments \cite{ssa1,ssa2,ssa3,ssa4}.

Almost all of our knowledge of the atomic scale structure of bulk condensed 
matter has 
been determined from measurements of the quantum mechanical interference
patterns that arise from particle-crystal scattering.  How do the new x-ray
holographies compare 
with crystallography, and what are the other possibilities? 
There are four equivalence classes of quantum mechanical interference patterns
that have been used to determine structure:
(1) In crystallography, there is an external source of particles which are sent
into the crystal in nearly plane wave states.
In the usual kinematic scattering limit, these particles coherently
single scatter from many atoms in the crystal. The interference between these
many single-scattering events produces the Bragg peaks \cite{dynamical}.
(2) In internal source holography, there is an internal source of particles
which leave the crystal in nearly spherical wave states.
These particles coherently single scatter from the object atoms in the crystal.
The interference between each of these single-scattering events and the strong
direct path reference beam produces the Gabor zone plates in the hologram.
(3) In external source holography, there is a coherent external source of 
particles which is sent into the crystal in nearly plane wave states.
These particles coherently single scatter from the object atoms in the crystal.
The interference between each of these single-scattering events and the strong
reference beam produces the hologram. Unfortunately, the necessary coherent
hard x-ray sources are not yet available \cite{slac} and when they become
available they will probably destroy the sample  in the process of making the
hologram \cite{trammel}.   (4) In the Kichuchi and Kossel methods, there is an
internal source of particles which leave the crystal in nearly spherical wave
states. These particles coherently multiply scatter from many atoms in the
crystal. The interference between these many multiple-scattering events
produces the Kichuchi and Kossel patterns. These multiple-scattering patterns
also contain useful holographic-like information \cite{usefulholo}, but this
information is different than the single-scattering holograms.

The remainder of this paper is organized as follows.
Sect.~II outlines the standard classical scalar wave derivation of the
intensity of the holographic interference pattern for the
interference between a spherical reference wave and a spherical
object wave.
Sect.~III is devoted to deriving an  expression for the corresponding
cross section for the intensity of the holographic interference pattern
for bremsstrahlung holography.
Since   this paper is concerned with 
the possible effects of virtual photons, it is sufficient to 
consider the process as being bremsstrahlung production by the source atom
followed by  photon scattering by the object atom, located at a 
displacement $\vec r$ from the source atom. 
We find that we can simplify the expression for this cross section and
apply it to holography if the atoms can be treated as well
separated so that only real photons propagate from the source atom to the
object atom.
We then show for real atoms in real crystals that the regions of high
electron density are sufficiently well separated. 
Our separated atom approximation is presented in Sect.~IV. 
The bremsstrahlung energies for the experiments we are considering are
40-60 keV and
at these energies, the x-ray--atom scattering cross section is dominated
by the Thompson process. So we study the photon virtuality effects for  
bremstrahlung production followed by Thomson scattering first.
For this case, 
the corrections to our separated atom approximation are defined and 
shown to be entirely negligible in Sect.~V.
  Near resonance, the anomalous 
scattering amplitude can become comparable to the
Thompson scattering 
amplitude, and these two amplitudes interfere.  We consider
this case in Sect.~VI, where we use the numerical  results of the 
previous sections to justify
 the immediate use of the separated atom approximation.
  In XFH, the photon is produced by fluorescence radiation,
where the excited  atomic state is produced 
by electron or photon impact, and the atom decays via photon emission.
The emitted photon can be scattered by another atom to 
produce an object
 amplitude which will interfere with the direct reference amplitude.
This is discussed in Sect.~VII, where the necessary 
amplitudes for this process are  presented.
  In MEXH, a real photon is sent into the sample from outside.  This photon has
a direct path to the detector atom and a collection of single scattering paths
to the detector atom
via the object atoms
 which will interfere with the direct path.  This is discussed 
in Sect.~VIII. The final Section is devoted to a brief summary and discussion.

\section{Classical Internal Source Ensemble X-Ray Holography}

When an x-ray photon is created inside a solid and is detected outside,
the quantum mechanical interference between the different
paths that 
the photon takes as it leaves the solid will produce a holographic image
of the atoms around the position where the photon is created.  
As we show below, the probability distribution for the photon intensity
is a Gabor hologram. 
In contrast to the
usual external source x-ray holography where 
the reference wave comes from outside
the sample, in internal source x-ray holography, the wave 
corresponding to the direct amplitude (i.e., the amplitude 
for the photon to leave the solid without any interactions)
serves as the
reference wave, and the wave corresponding to the amplitude produced by
single photon--atom scattering plays the role of
the object wave.  

Because the amplitude for photon--atom scattering is weak for hard x-rays, the
reference wave will be much  stronger than the object waves, and this  strong
reference wave limit is the ideal holographic situation because the hologram is
then dominated by the interference between the reference wave and the singly
scattered object waves.  In this limit, the interference between one object
wave and another object wave, and the interference between the reference wave
and the low order (double, triple, ...) multiple scattering object waves is
much weaker than the interference between the reference wave and the
single-scattering object waves.  In nearly perfect crystals, the interference
between the high order object waves can become comparable to the reference wave
and is responsible for the Kossel (x-ray) and Kikuchi (electron) patterns. 
However, these multiple scattering features are sharp in angle and therefore
can be easily removed from the hologram.

To develop the simple classical  wave picture for internal source ensemble
x-ray holography, consider first just the two atoms shown in Fig.~1.  The full
internal source x-ray hologram can be obtained from this two atom case  by
summing over all object atoms for each source atom, and by summing over all
source atoms. Suppose the source atom at the origin emits radiation which is
detected in the far field, and suppose also that prior to detection the
radiation is scattered by a second object atom located at position $\vec a$. 
The direct and single-scattering  paths produce an interference pattern.  When
the polarization is not  important, this problem can be  treated as due to the
interference between two scalar  wave fields, with the scalar field
representing  a component of $\vec E $ or $\vec B$.

In this approximation, the first atom emits a scalar spherical 
reference wave $R$ of the form 
\begin{equation}
R = {e^{ik r} \over r}, 
\label{req}
\end{equation}
and the second atom emits a scalar spherical object wave $O$ of the 
form
\begin{equation}
 O = {e^{ika}\over a} \; f_a(\theta) \; 
{e^{ik |\vec r -\vec a|}\over|\vec r -\vec a|},
\end{equation}
 where again $\vec a$ 
is the position of the  second atom. 
In the far field, $r\gg a$,  the 
composite amplitude $M = R+O$ takes 
the form 
\begin{equation}
M = {e^{ik r}\over r}
(1+e^{ika} \; {f_a(\theta) \over a} \; e^{-i\vec k\cdot\vec a}) 
\label{meq}
\end{equation}
where $f_a(\theta)$ is the standard atomic scattering amplitude for
an incident plane wave, and we have followed the tradition in this
field of neglecting the higher-order spherical wave corrections.

The far field intensity is given by the square of the composite amplitude $M$
\begin{equation}
I = M^{*} M = R^{*} R + R^{*} O + R O^{*} + O^{*} O .
\end{equation}
Holography records much more phase information than crystallography,
but there is still a ``holographic phase problem" present in the holographies
that can only
measure the intensity 
({\it{e.g.,}} laser, electron and x-ray holography) and not the
amplitude  ({\it{e.g.,}} acoustic and microwave holography).
The intensity holographies 
record both the information we want about the object in the 
$R^{*} O$ term, and a copy of the complex conjugate of this information 
in the $R O^{*}$ term, 
which produces a non-existent twin to the object during the
reconstruction.

The far field object plus twin holographic interference cross term 
$R^{*} O + R O^{*}$
is proportional to
\begin{equation} 
Re\left[ f_a(\theta) \right] cos (ka-\vec k\cdot\vec a)
-Im\left[ f_a(\theta) \right] sin (ka-\vec k\cdot\vec a).
\end{equation}
In this paper, we treat the 
case in which the source atom emits x-rays with wavelength $\lambda$.
Note that to obtain significant holographic oscillations, it is important
to make $\lambda \le a$, and that the best holograms will be produced when
$\lambda \ll a$.

Although all of the existing and proposed 
internal source holographic techniques actually depend on the
multi-path quantum mechanical
interference of the particle emitted by the sample,
the essential features of the holograms can be (and have been before this paper)
obtained from 
the simple wave picture of the process outlined above.

There are two 
essential ingredients of atomic resolution internal source holography:
(1) There is a 
localized source inside the sample.  This localization provides the necessary
spatial coherence of the source.  The particle can be localized by being
created inside the sample---this is the case for the bremsstrahlung and
fluorescence x-ray holographies described in this paper.  
The particle can also be localized by 
being ejected from a specific quantum state in the sample---this is the case for
photoelectron and Auger electron holography.  
In addition, the particle can be localized
by an incoherent inelastic  scattering event---this is the case for
diffuse LEED holography and diffuse Kikuchi electron holography. 
The analogous incoherent scattering localization
is possible theoretically for photons via thermal diffuse x-ray
scattering and via Compton scattering.  
(2) There is interference between the
direct reference wave and the singly scattered object waves.  This requires 
a coherent scattering event at the object atoms.  If the object atoms
scatter incoherently, the interference in the final state will not occur.

We shall show how these features arise from quantum electrodynamics.

\section{Bremsstrahlung X-Ray Holography (BXH)}

In the bremsstrahlung process, an electron incident on a solid 
radiates a photon:
$e(p_i)+{\rm solid}\to e'(p_f)+\gamma(k)+{\rm solid}'$.
 excited. The effects of thermal motion are ignored.
In this paper, we consider the case where the initial and final electronic
states of the atoms are the same, and where the solid is a collection of
fixed atoms, {\it{i.e.,}} we do not consider the effects of thermal motion.
The Feynman diagrams for bremsstrahlung holography are shown in Figs.~2 and 3.
The complete holographic amplitude ${\cal M}$ is the sum of the reference wave 
amplitude ${\cal R}$,
given by the crossed and
 uncrossed Born bremsstrahlung terms shown in Fig.~2, and the
object wave amplitude
${\cal O}$, given by the
 crossed and uncrossed Compton scattering terms shown in Fig.~3:
\begin{equation}
{\cal M} ={\cal R}+{\cal O}.
\end{equation}
Here the QED reference ${\cal R}$, object ${\cal O}$, and hologram ${\cal M}$ 
amplitudes 
are analogous to the corresponding classical $R$, $O$, and $M$ terms in
Eqs.~(\ref{req}) through (\ref{meq}).

Before starting, 
it is useful to sketch the notational conventions used in this paper.
The various four-momentum 
vectors are represented with Roman typeface, {\it{e.g.,}}  k, p$_i$, p$_f$,
and the three-vector spatial components are indicated with italic typeface
with explicit arrows, {\it{e.g.,}}
$\vec k$, $\vec p_{i}$, and $\vec p_{f}$.
The magnitudes of three-vectors are written explicitly, {\it{e.g.,}} $|\vec k|$.
Thus for the bremsstrahlung reference amplitude shown in Fig. 2, the
conservation of energy and 
momentum is written as ${\rm{p}}_i + q = {\rm{p}}_f + k$,
where $\omega = {\rm{k}}^0$,  
${\rm{p}}_{i}^0 = E_{i}$, and ${\rm{p}}_{f}^0 = E_{f}$. 
Here q is the four-momentum supplied by the target, q = (0,$\vec q$),
where the $0$ arises from our condition that
no atoms (or  nuclei) be excited.
Three-momentum conservation is expressed as $\vec q+\vec p_i=\vec p_f+\vec k$.
The notation and conventions of Bjorken and Drell \cite{bd}
are used throughout the present work, and our units are such that 
both $\hbar$ and $c$ are unity.

The cross section for the holographic interference pattern
is related to the square of the holographic amplitude ${\cal M}$ by:
\begin{equation}
{d^3\sigma \over d\Omega \; d\Omega_f \; d | \vec k |} 
= m^2 \; {p_f\over p_i} \; 2\pi \;
{\omega \; \overline{|{\cal M}|^2} \over 2 \; (2 \pi)^6}
\; \theta(E_i-m-\omega) 
\label{e:one}
\end{equation}
where $m$ is the mass of the electron,
  $\Omega$ represents the 
outgoing
angles of the photon and $\Omega_f$ those of
the electron $(p_f)$.  The quantity
$\overline{|{\cal M}^2|}$ is obtained from 
$|{\cal M}|^2$ by squaring the magnitude of 
${\cal R}+{\cal O}$, summing
over the spins of the final electron, and averaging over the spins of the 
initial electron.

The reference amplitude {\cal R} shown 
diagrammatically in Fig. 2 is evaluated as
\begin{equation}
{\cal R}(k,q)={Ze_pe^2\over |\vec q\;|^2}
\left[\bar u_f\left\{\left({\epsilon\cdot 
p_f\over 2p_f\cdot k} - {\epsilon\cdot p_i\over 2p_i\cdot k}\right)\gamma^0+
{\rlap/\epsilon\rlap/k\gamma^0\over 2p_f\cdot k}+{\gamma^0\rlap/k\rlap/\epsilon
\over 2p_i\cdot k}\right\}u_i\right]\left(1-F(|\vec q \; |)\right)\label{e:two}
\end{equation}
where $e^2/4\pi =\alpha$, the proton charge is the negative of the electron 
charge  $e_p = -e$, and $F(|\vec q \; |)$ is the electronic contribution to the 
atomic form factor, normalized so that $F(0)=1$. 
%
This matrix element is proportional to $\epsilon_{\mu} = (0,{\hat \epsilon})$.
 The nuclear form factor is essentially unity for the 
kinematic range of the current x-ray holography experiments.
The deviation of the term $\left(1-F(|\vec q \; |)\right)$ from unity
represents the  screening effect of  the atomic electrons. If $\vec q=0$, the
atom acts as a neutral object and there is no bremsstrahlung. 

The only significant approximation made in 
obtaining Eq.~(\ref{e:two})  is that
the initial and final state electron--nuclear Coulomb interactions
have been neglected. The influence of these
interactions, which can increase the value of the computed cross 
sections significantly, can be reasonably well approximated by multiplying the 
above amplitude
 by the product of the continuum electronic wave functions evaluated 
at the nuclear center---this is the Elwert approximation of Ref.~\cite{elwart}. 
This is a well motivated approximation classically because the
 acceleration that 
leads to the bremsstrahlung takes place in the vicinity of the nucleus.
Detailed numerical 
studies \cite{pratt} have confirmed the qualitative accuracy of the 
Elwart approximation.  Multiplying our  amplitude by this factor 
does not influence the propagation of the virtual photon 
between atoms, which is 
our principle concern. Thus we shall ignore
the initial and final state interactions here in our study of the
potential off-shell effects.

It is convenient to define the expressions in the bracket as $\epsilon\cdot B
(k)$, so that the reference amplitude can be rewritten as
\begin{equation}
{\cal R}(k,q) = {Ze_pe^2\over |\vec q\;|^2}\;(1-F(|\vec q \; |)) \;
B_\mu(k)\epsilon^\mu.
\label{e:three}
\end{equation}

As shown in Fig.~3, 
the virtual photon ($k'$) is produced by the source atom,
propagates to the object atom, located at a 
separation $\vec r$ from the source, which scatters the virtual photon $k'$
so that the final photon $k$ is produced. 
The object atom scattering is dominated by the 
Compton scattering of the
photon by the atomic electrons. This is because
the photon--atom scattering is larger than  
the photon--nuclear scattering, by the ratio
of the proton mass to the electron mass for the Thompson term, or by 
the ratio of  the squares of the atomic and nuclear radii for the dipole 
terms.  The virtual 
bremsstrahlung matrix element is denoted as ${\tilde B_\mu(k')}$
and the Compton rescattering  transition 
matrix as C$^\mu(\epsilon,k, k')$. The evaluation of the Feynman graphs 
shown in 
Fig.~3 uses  standard techniques\cite{bd}. 
Here we also carry out the integration over the time 
component of $k'$, which gives us a delta function setting $k'^{\;0}=\omega$. 
Then we arrive at the expression for the object amplitude:
\begin{equation}
{\cal O} = -Z^2 e^4 e_p\int <C^\mu(\epsilon,k,k')> {\tilde B_\mu
(k')
\over [\omega^2-\vec k'^{2}+i\epsilon]} 
e^{-i(\vec k-\vec k')
\cdot
\vec r}{d^3k'\over (2\pi)^3}{\left(1-F(|\vec q\; '|)\right)
\over \vec {q\;'}^2 +i\epsilon}
\label{e:four}
\end{equation}
where 
$\vec q\; '=\vec p_f - \vec p_i+\vec k'$.  
Note that 
$k'^{2} = \omega^2-\vec k\;'^{2}\ne 0$. 

The 
bremsstrahlung matrix element is 
given by 
\begin{equation}
\tilde B_\mu(k')\equiv \bar u_f\left\{\gamma_0{1\over\rlap/p_i-\rlap/k'+m}
\gamma_\mu+\gamma_\mu\;{1\over\rlap/p_f+\rlap/k'-m}\gamma_0\right\}u_i,
\label{e:six}
\end{equation}
where standard spinor notation is used.
Note that the Compton term $<C^\mu(\epsilon,k,k')>$ is the atomic expectation
value of  the virtual-to-real Compton transition matrix which converts the
virtual  photon k$\; '$ to a real one with four momentum and polarization
(k, $\epsilon$). Thus
\begin{equation}
C^\mu(\epsilon,k,k')=\gamma^\mu S_F(P_i-k)
\rlap/\epsilon +\rlap/\epsilon\; S_F(P_f+k)\gamma^\mu,
\label{e:fivea}
\end{equation}
where S$_F$(P) is the relevant propagator for the bound electrons.
For example, when the electrons are treated 
as free, the Compton transition matrix 
element
is given by 
\begin{equation}
C^\mu_{f,i}(\epsilon,k')\equiv
\bar U_f\left\{\gamma^\mu{1\over\rlap/P_i-\rlap/k-m}
\rlap/\epsilon +\rlap/\epsilon\;{1\over \rlap/P_f+\rlap/k-m}\gamma^\mu\right\}
U_i
\label{e:five}
\end{equation}
where
the upper case spinors $U_i$, $U_f$ represent the initial and final
states of the free electron,
with $\vec P_i +\vec k'=\vec P_f+\vec k$.

But the atomic bound states are more interesting. We may better understand
the operator of Eq.~(\ref{e:fivea}) by noting that
in the relativistic theory, 
 the origin of the 
Thompson term comes from the terms involving the creation of virtual
electron--anti-electron pairs \cite{Sakurai}.
The resulting two-electron plus anti-electron states live only
for a very short time, so that we may  ignore interactions with the other 
particles of the atom. The remaining terms can be seen 
in the non-relativistic limit up to ${\cal O}(p^2/m^2)$ as arising from the 
two interactions of the dipole operator\cite{Sakurai}.
Then, we may write 
 the Compton transition matrix element as a sum of terms so that 
\begin{eqnarray}
<C^\mu(\epsilon,k,k')>= <T^\mu(\epsilon,k,k')>+ <R^\mu(\epsilon,k,k')>
\end{eqnarray} 
where the 
Thompson scattering  is denoted as 
T$^\mu(\epsilon,k,k')$ and the resonant scattering 
and other contributions are denoted as $<R^\mu(\epsilon,k,k')>$. 

When the long wavelength approximation is 
valid, we obtain
\begin{eqnarray}
<R^\mu(\epsilon,k,k')>=
\omega^2e^2\sum_n {<i|\hat\epsilon\cdot \vec D|n><n| D_l|i>\over 
E_i+\omega-E_n+ {1 \over 2} i \Gamma_n}\delta_{\mu,l} + \nonumber \\ 
\omega^2e^2\sum_n {<i|  D_l |n><n|\hat\epsilon\cdot\vec D|i>\over
E_i-\omega-E_n}\delta_{\mu,l},
\label{e:rdef}
\end{eqnarray}
where the dipole operator $\vec D$ is given by 
\begin{eqnarray}
\vec D= \sum_{i=1}^Z \vec s_i
\label{e:jdef}
\end{eqnarray}
and $\vec s_i$ is the displacement of the
i--th electron from the atomic center. The vector $\vec D$ is simply the 
sum of electronic dipole operators. The quantities $E_n$ and $\Gamma_n$
are the energy and the width of the excited state n. 
The result shown in Eq.~(\ref{e:rdef}) indicates that only the three-vector  
$\vec R$ part of 
$R^\mu$ enters into the expression for the amplitude. 
This is because each atomic photon emission/absorption is controlled by
an $\hat\epsilon\cdot\vec D$ operator.
For 
unpolarized atoms, $\vec R$ must be proportional to 
$\hat\epsilon$, so that it is convenient to define a strength 
function $S(\omega)$ such that:
\begin{equation}
\vec R=\hat\epsilon \; S(\omega).
\label{e:rsimp}
\end{equation}
We use this definition along with Eq.~(\ref{e:rdef})
to obtain 
\begin{equation}
 S(\omega)=\omega^2e^2\sum_n |<i|\epsilon\cdot \vec D|n>|^2\left[
{1\over E_i+\omega-E_n+{1 \over 2} i \Gamma_n} + 
{1\over E_i-\omega-E_n}
\right].
\label{e:sdef}
\end{equation}

As noted in the introduction, we shall proceed by first studying the 
effects of Thompson scattering by the object atoms. 
An explicit evaluation yields
\begin{equation}
<T^\mu> = -{1\over m}\;\delta^{\mu i}\;\epsilon_i \; F(|\vec k -\vec k'|),
\label{e:eight}
\end{equation}
where for simplicity
we take the scattering object atom to be of the same type as the 
source atom which 
produced the virtual photon.
In this and the following two sections we shall consider 
photons 
for which the Thompson  term is dominant. We shall return to the 
resonant corrections to the Thompson term in Sect.~V.

We compute $\overline{|{\cal M}^2|}$ by squaring ${\cal R}+{\cal O}$, keeping 
only the Thompson scattering contribution $T^\mu$ in ${\cal O}$, 
by summing over the final electron spin, 
and by averaging over the initial electron spin.  
The result is 
\begin{eqnarray}
\overline{|{\cal M}^2|} 
& = & 
       {1\over 2}\;\sum_{s_fs_i}\biggl[B^*\cdot\epsilon \;B\cdot
       \epsilon\biggl({Ze^3\over |\vec q\;|^2}\;(1-F(|\vec q\;|))\biggr)^2
+ {Z^3e^8\over |\vec q\;|^2}\;(1-F(|\vec q\;|)) 
\nonumber \\
& \times & 
       2\; Re\biggl\{B^*(k)\cdot
       \epsilon\int {d^3k'\over (2\pi)^3}\;
       {(1-F(|\vec q\; '|)\over |\vec q\; '|^2}
       \; e^{-i(\vec k-\vec k')\cdot\vec r}
       \; {F(|\vec k-\vec k'|)\over
       \omega^2-\vec k'^{2}+i\epsilon}
\nonumber \\
& \times &
        < T^\mu(\epsilon,k,k')>\tilde B_\mu(k')\biggr\}\biggr].
\label{e:seven}
\end{eqnarray}
Here the
 term of order $Z^4e^{10}$, which is much smaller than the lower order terms, 
has been neglected.

The result given by Eq.~(\ref{e:seven}), as specified by the matrix elements
given by  Eqs.~(\ref{e:six}) and (\ref{e:eight}), is our main result. It gives
the bremsstrahlung holography cross section when the object
scattering is dominated by 
the Thompson term. 
We shall make a further simplification to facilitate a
first evaluation: we will keep only the numerically most significant 
terms of $B$ and $\tilde B$, {\it{i.e.,}} the ones proportional to 
$\hat \epsilon\cdot \vec p_f$ and 
$\hat\epsilon\cdot \vec p_i$. This leads to the 
standard classical expression  
for the bremsstrahlung cross section \cite{Jackson}.
We have explicitly evaluated the neglected terms numerically and found
that their neglect   
produces an approximately constant 10\% reduction 
in the cross section over the kinematic region where the 
bremsstrahlung holography experiments will operate.  
Our numerical results can be understood by noting that
$\rlap/\epsilon\rlap/k\gamma^0\approx i\vec\sigma\cdot\vec\epsilon\times
\vec k$. This spin-dependent interaction is a magnetic effect
proportional to 
$\vec \nabla \times \vec A$, which therefore does not 
interfere with the terms we keep. Furthermore, the two 
spin-dependent terms of Eq.~(\ref{e:two}) partially 
cancel and their sum is smaller than the leading term by about $k/m$.

We carry out the average over electron initial spin, and the 
sum over final electron spin.
The result for the bremsstrahlung holography cross section is: 
\begin{equation}
{d^3\sigma\over d\Omega \; d\Omega_f \; d|\vec k|} =
m^2
\; {|\vec{p_f}|\over|\vec{p_i}|}
\; {1\over 2}
\; {\omega\over (2\pi)^5}
\; \theta(E_i-m-\omega)
\; \overline{|{\cal M}|^2}
\; {1 \over 8m^2}
\; [8E_iE_f- 4 p_i\cdot p_f + 4m^2]
\label{e:nine}
\end{equation}
where
\begin{eqnarray}
\overline{|{\cal M}|^2} 
& \equiv & 
\left({Ze^3\over |\vec q\;|^2}\left(1-F(|\vec p_f -
\vec p_i + \vec k|)\right)\right)^2 
\; [\hat\epsilon \cdot \vec {V}(k)
\;  \hat\epsilon\cdot\vec{ V}(k)] 
{Z^3e^8\over |\vec q\;|^2}\;{2(1-F(|\vec q \; |))\over m}\; 
Re \; I(\vec k,\vec r) \; .
\label{e:ten}
\end{eqnarray}
Here the quantity $I(\vec k,\vec r)$ is 
given by
\begin{eqnarray}
I(\vec k,\vec r) \equiv
\hat\epsilon \cdot \vec {V}(k)
\; \int {d^3k'\over (2\pi)^3}
\; \hat\epsilon\cdot\vec{ V}_1(k')
\; e^{-i(\vec k-\vec k')\cdot\vec r} 
\; \biggl\{{\left(1-F(|\vec q\; '|)\right)\over |\vec q\; '|^2}\;{F\left(|\vec k
-\vec k'|\right)\over\omega^2-\vec k'^{2} + i\epsilon}\biggr\}\,\;,
\label{e:iint}
\end{eqnarray}
where $\vec q\;'
=\vec p_f-\vec p_i+\vec k'$ and $\vec q = \vec p_f - \vec p_i+\vec
k$.  The vectors $\vec {V}(k)$ and $\vec {V}_1(k')$ are given by
\begin{equation}
\vec {V}(k) \equiv {\vec{p_f}\over p_f\cdot k} - {\vec{p_i}\over p_i\cdot k} 
\label{e:eleven}
\end{equation}
\begin{equation}
\vec{V_1}(k') \equiv 
{2\vec{p_f}\over 2p_f\cdot k'+k'^{2}} - {2 \vec{p_i}\over 2p_i\cdot k'-k'^{2}}.
\label{e:twelve}
\end{equation}

The cross section for the intensity of a bremsstrahlung hologram given
by Eq.~(\ref{e:nine}) together with the definitions given by 
Eqs.~(\ref{e:ten}) through~(\ref{e:twelve}) is the complete solution
to the bremsstrahlung holography problem.  What remains to be done is to
carefully analyze these equations to see how the classical holography
equations emerge in the classical limit, and to see how large the quantum
effects are, and when they are important.  That is the content of
the next three sections.

\section{ Separated Atom Approximation}

The goal of the bremsstrahlung holography experiments 
is to determine precise information about the 
location of the object atoms, which is represented 
in Eqs. \ (\ref{e:nine}) through (\ref{e:twelve})
by the vector $\vec r$.
The standard holography expressions involve an interference term of the
general 
form $\beta \; e^{i\omega r} e^{-i\vec k \cdot \vec r} / r$ where $\beta$ is 
a known function. A quick look at Eqs.\ (\ref{e:ten}) and (\ref{e:iint})
could lead one to dismay. How could that integral ever have the simple
form required for holographic investigations?
We indicate a solution
by considering the situation when 
the two atoms are very far apart, {\it{i.e.,}} in the limit where r approaches 
infinity.
Here our intuition provides a  guide:
the process must proceed by  bremsstrahlung from the source atom 
followed by photon propagation 
along the direction of $\vec r$ and Thompson scattering by the object atom.
Thus the bremsstrahlung 
makes a real photon with momentum $\vec \kappa\equiv \omega \hat r$
and energy $\omega$,
and the Thompson scattering changes the direction of $\vec\kappa$ to 
$\vec k$. 

In this case, the photon has four momentum $\kappa\equiv(\omega,\vec \kappa)
$ and $\kappa\cdot\kappa =0$. The propagating photon is on-shell.
This situation is simple, but the full integral of Eq.~(\ref{e:iint}) is not.
 However, we will evaluate this integral 
by developing expansions in which the leading 
term is correct in the limit that r is very large. We shall keep the 
leading term and the most important corrections. We call this approach 
the separated atom approximation. In practice, this
amounts to replacing $k\; '$ by $\kappa$ in certain terms in 
the integrand. 
For example, we shall show below that replacing 
$\vec V(k\; ')$  by $\vec V(\kappa)$ and $\vec q\; '$ by
$\vec p_f-\vec p_i+\vec \kappa$ are excellent approximations.

We may use the uncertainty principle to better understand why the 
propagating photon must be real for infinite values of r.
Our  virtual photons 
have energy $\omega$, but the magnitude of the three momentum 
varies from $0$ to infinity in the integration. Let us define 
$Q^2\equiv \omega^2 -|\vec k\; '|^2$ to provide a 
measure of the violation of 
energy conservation
required to make the virtual photon. This is simply the square of the 
energy-momentum four vector, which vanishes for real photons.
If $Q^2<0$ the wave is a decaying exponential of the form
$e^{-|\hbar Q r|}/r$, 
which has a small value. The interpretation of this
Yukawa form is that the  
photon lives for a time
$\hbar/ Q $, so that its maximum range is $\hbar c/ Q $.
If $Q^2> 0$, the wave is of the form $e^{i \hbar Q r}/r$. The effect
of this term is very small because of the oscillations of the 
integrand in the integral over $d^3k\; '$.
Thus, the net result is that only
the real photons with $Q=0$ reach the object atom.

Are the atoms in a real solid sufficiently separated so that all the
virtual photon effects are gone before the bremsstrahlung photon reaches
the nearest atoms? 
We argue that the answer is yes, 
at least for most solids at typical experimental bremsstrahlung 
holography energies, 
by considering the specific example of crystalline copper.
In crystalline copper the atoms are separated by a distance 
$R_N \approx 2.5561 \; {\rm{\AA}} \approx 4.83 a_0$  where $a_0$ is 
the Bohr radius $\sim 0.529 \; \rm{\AA}$. Where are the electrons in each atom?
The electron density for isolated copper atoms and for
crystalline copper calculated using the FEFF computer code \cite{Rehr} 
is shown in  Fig.~4.  Note 
that this density is sharply peaked at small values of $s$
since most of the electrons 
are within $1 \; \rm{\AA}$ of the nuclear center of the atom, and that the
electron densities for 
isolated atoms and for atoms embedded in the solid are very 
similar.
In particular, the root mean square 
radius of the displayed density is 1.08 $a_0$.
Thus the closest separation $r$ between the copper atoms
is about 5 times the typical value of the distance $s$ between an 
electron and the nucleus.
At the very least, it is reasonable to 
 expect that the separated atom approximation is a good starting point.

Our procedure is to 
examine a set of approximations to the full results for 
$I(\vec k,\vec r)$ given by Eq.~(\ref{e:iint}).
We will define the on-shell, separated atom 
approximation as the result of
setting $\vec k'=\vec\kappa$ in $V_1( k')$ 
and in $\vec q\; '$.  Then the on-shell approximation $I_{on}(\vec k,\vec r)$
to the full on- and off-shell integral $I(\vec k,\vec r)$ is given by
\begin{equation}
I_{on}(\vec k,\vec r) = {\left(1-F(|\vec p_f - \vec p_i +\vec \kappa|)\right)
\left(1- F(|\vec q \; |)\right)\over |\vec p_f - \vec p_i+\vec \kappa\;|^2} 
\; \hat\epsilon\cdot \vec V(k) 
\; \hat\epsilon \cdot \vec V(\kappa)
\; J_{on}(\vec k, \vec r),
\label{e:fourteen}
\end{equation}
where 
\begin{equation}
J_{on}(\vec k, \vec r)= \int
{d^3k'\over (2\pi)^3} 
\; e^{-i(\vec k-\vec k')\cdot \vec r} 
\; {F(|\vec k - \vec k'|)\over \omega^2 -\vec k'^{2} + i\epsilon}.
\label{e:fourteena}
\end{equation}
Again, the subscript ``$on$" 
is to remind us that
setting $\vec k'$ equal to $\vec\kappa$ causes the propagating photon to be 
on-shell. Its virtuality has decayed by the time it reaches the near-neighbor
atoms, and the square of its four momentum has vanished.
The next section (Sect.~V) is devoted to the demonstration that 
the on-shell $I_{on}(\vec k, \vec r)$
given by Eq.~(\ref{e:fourteen}) is an excellent 
approximation to the full
$I(\vec k, \vec r)$ given by  Eq.~(\ref{e:iint}).

 The first step is to 
understand the integral $J_{on}$.
We can gain some insight 
 by converting the integral over the momentum into one 
involving positions. We use 
\begin{equation}
F(|\vec k -\vec k'|) \equiv \int d^3s\;\rho(s) e^{-i(\vec k-\vec k')\cdot
\vec s},
\label{e:fifteen}
\end{equation}
so that the on-shell $J_{on}$ integral given in Eq.~ (\ref{e:fourteena}) 
simplifies to 
\begin{equation}
J_{on}(\vec k,\vec r) = -{1\over 4\pi}\int d^3s\;\rho(s)\;{e^{i\omega
|\vec s+\vec r|}\over |\vec s+\vec r|} e^{-i\vec k\cdot (\vec s+\vec r)}.
\label{e:sixteen}
\end{equation}
If $r\gg s$ for the important regions of $\rho(s)$, we may 
replace ${e^{i\omega
|\vec s+\vec r|} / |\vec s+\vec r|}$ by ${e^{i\omega r} }
 e^{i \omega\vec r\cdot \vec s} / r$. 
This gives
\begin{equation}
\lim_{r\to\infty} J_{on}(\vec k,\vec r) = -{1\over 4\pi}
\;{e^{i\omega r} \over r}
\; e^{-i\vec k\cdot\vec r}
\; F( | \vec k - \vec \kappa |),
\label{e:seventeen}
\end{equation}
which has the usual classical holographic form. 
The spherical Green's function ${e^{i\omega r} / r}$ corresponds to 
the form of the wave at infinity.  
This form arises from the pole in the integral for $J_{on}$
at $|\vec k'|=\omega$.

 To check the asymptotic approximation 
given in Eq.~(\ref{e:seventeen})
we numerically compared the full $J_{on}$ given by Eq.~(\ref{e:sixteen})
with its approximation given by Eq.~(\ref{e:seventeen}).
The results are shown in Fig.~5. Note that the approximation is excellent 
except when  $\vec k \parallel \vec r$.  Even then,
the full theory and the approximation
produce very similar 
holographic interference patterns; the asymptotic approximation
produces a pattern about 10 \%  smaller than the full theory when 
$\vec k \parallel \vec r$.
This agreement between the full quantum electrodynamic calculation and
the simple classical holography equations shows
that bremsstrahlung holography is possible: the reduction for 
$\vec k \parallel \vec r$ does not 
significantly change the oscillatory form with its 
strong dependence on $\vec k\cdot \vec r$.

There is no need to use the asymptotic 
approximation in numerical work. We may use the correct value of $J_{on}$
and maintain the explicit holographic form.
This involves expanding the form factor in terms of Legendre 
polynomials $P_L(\hat k \cdot \hat k')$:
\begin{eqnarray}
F(|\vec k -\vec k'|)=\sum_L F_L(\omega r)P_L(\hat k \cdot \hat k'),
\label{e:partialwaveF}
\end{eqnarray}
where $\omega =|\vec k|$. 
Combining the partial wave expansion for 
the atomic form factor
given by Eq.~(\ref{e:partialwaveF}) 
with the full expression for $J_{on}$ given by Eq.~(\ref{e:fourteena}) yields
the partial wave expression for $J_{on}$
\begin{eqnarray}
J_{on}(\vec k,\vec r) =i\omega \sum_L i^L F_L(\omega r) h_L^{(1)}(\omega r)
P_L(\hat k \cdot \hat r),
\label{e:exact}
\end{eqnarray}
where $h_L^{(1)}(\omega r)$ are the outgoing spherical Bessel functions.
All that is required for this to hold 
is that $r$ be bigger than the 
maximum value of $s$ ({\it{i.e., circa}}~ $2.56 \; \rm{\AA}$ 
for copper) occuring in the
integral~(\ref{e:sixteen}). 
Since $r= 2.56 \; \rm{\AA}$ and Fig.~4 shows that $\rho(s)$ is less than 
$10^{-3}$ of its maximum value for $s \ge 0.5~\rm{\AA}$,
this condition is met.
We may also understand the relation between this expression
and its limiting form shown in Eq.~(\ref{e:seventeen}). The use of the 
asymptotic form of the outgoing spherical Bessel functions:
\begin{eqnarray}
\lim_{x\to\infty}h_L^{(1)}(x)=(-i)^{L+1}{e^{ix}\over x},
\label{e:asympbessel}
\end{eqnarray}
leads immediately to the result shown in Eq.~(\ref{e:seventeen}).
The corrections to this asymptotic form are thus of order $1/x$ times 
the original result. Thus we see that the expected first correction to 
Eq.~(\ref{e:seventeen}) is of the order of $1/\omega r \approx 1/25$ for 
$\omega= 20$ keV. 

Equation~(\ref{e:exact}) allows us to understand why the difference between the 
asymptotic approximation given by Eq.~(\ref{e:seventeen}) and the exact result
given by Eq.~(\ref{e:exact}) is largest for $\vec k\cdot\vec r = 1$.  The
terms $F_L (\omega r) h_L (\omega r)$ monotonically approach zero as $L$ 
increases.  The function 
$P_L(\vec k\cdot\vec r = 1) = 1$ for all $L$, so that the
terms with large $L$ (these are the terms for which the approximation 
generated by Eq.~(\ref{e:asympbessel}) is less accurate) add constructively.
This can also be 
seen (without using the partial wave expansion) by examining the
integrand of Eq.~(\ref{e:sixteen}).  If $\hat k\cdot\hat r = 1$ ({\it{i.e.,}}
when $\vec k \parallel \vec r$), the term 
$\omega |\vec s + \vec r | - \vec k \cdot \vec s $ 
is greatly reduced so that the contributions of the larger values of s are 
less inhibited by the oscillating exponential than for other values of
$\vec k\cdot\vec r$.

The partial wave 
expression (\ref{e:exact}) systematically gives all of the corrections 
to the classical holographic form given by
Eq.~(\ref{e:seventeen}). However, it is useful to provide another
approximation which shows us why the relevant integrals are dominated by
terms in which $\vec k^{\; '}= \vec \kappa$.
The idea is to approximate $F(|\vec k -\vec k'|)$ using
\begin{eqnarray} 
F(|\vec k -\vec k'|)\approx
F(q_1) +(\vec k'-\vec\kappa)\cdot \vec{\nabla}_{q_1} 
F(q_1)\label{e:expand}
\end{eqnarray}
where $q_1\equiv|\vec k-\vec\kappa|$.
We use (\ref{e:expand}) in the integral (\ref{e:fourteena}) and note that 
 the $\vec k'$ appearing in the numerator 
of the integral (\ref{e:fourteen}) can be replaced by a gradient on $\vec r$.
Thus we find
\begin{eqnarray}
J_{on}(\vec k, \vec r) \approx 
-{1\over 4\pi}\;{e^{i\omega r}
\over r} e^{-i\vec k\cdot\vec r} \; F(q_1)
+ \delta J_{on},
\end{eqnarray}
with 
\begin{eqnarray}
\delta J_{on}(\vec k, \vec r) \equiv 
-e^{-i\vec k\cdot\vec r}
\vec\nabla_{q_1}F(q_1)\cdot \vec V_{on}(\vec k,\vec r),
\end{eqnarray}
where
\begin{eqnarray}
\vec V_{on}(\vec k,\vec r)\equiv\left({\vec\nabla_r \over i}-
\vec\kappa\right)
\int 
\; {d^3k'\over (2\pi)^3} 
\; e^{i\vec k'\cdot\vec r}
\; 
\biggl\{{F(|\vec k -\vec k'|)\over 
\omega^2-\vec k^{\;'^2} + i\epsilon}\biggr\} \; . 
\label{e:vondef}
\end{eqnarray}

One may use the Legendre partial wave expansion of Eq.~(\ref{e:exact}) 
above to obtain a more 
detailed expression for $\vec V_{on}(\vec k,\vec r)$. But the main point is 
that the long distance behavior
of the integral is that of 
$e^{i\omega r}$.  Since $(-i {\vec\nabla_r} -
\vec\kappa)
e^{i\omega r} =0$, the correction to the separated atom approximation
must have an extra factor of order $1/\omega r\approx 1/25$.
We shall use expansions similar to that of Eq.~(\ref{e:expand}) to 
systematically understand the short distance terms. The integral 
$\vec V_{on}(\vec k,\vec r)$ will appear again. Furthermore, we shall often 
employ the technique of writing a complete expression as its on-shell 
approximation plus a term which 
is proportional to $\left(-i{\vec\nabla_r}-
\vec\kappa\right)$ and 
vanishes in the asymptotic limit
given by Eq.~(\ref{e:seventeen}).

\section{Short range terms}
In the previous section we showed how keeping the effects of the pole
at $|\vec k'|=\omega$ led to the term with the long distance propagation.
Here we show that this pole dominates the complete expression given by Eq.~(\ref
{e:iint}). 
The vector $\vec k'$ appears in three places in this integral,
in $\left(1-F(|\vec q\; '|)\right)$, 
in $1/|\vec q{\;'}|^2$, and in $\vec V_1(k')$.
We will denote these terms as the screening correction, the Coulomb photon 
propagation, and the electron propagation. We shall study each one separately.

\subsection{Screening correction}
Keeping the $\vec k'$ in the screening term leads to the integral, $I_s$:
\begin{equation}
I_s(\vec k,\vec r) = \int {d^3k'\over (2\pi)^3} 
e^{-i(\vec k - \vec k')\cdot\vec r}
\biggl\{\left(1-F(|\vec q\; '|)\right)
{F(|\vec k -\vec k'|)\over 
\omega^2-\vec k'^{2} + i\epsilon}\biggr\}.
\label{e:is}
\end{equation}
Recall that $\vec q\; '=\vec p_f-\vec p_i +\vec k'$. 
Pole dominance of the integral would allow us to replace the $\vec k'$ 
appearing in $\vec q\; '$ by $\vec\kappa$. Thus the integral may be 
 approximated by using
\begin{eqnarray}
 F(|\vec q\; '|)=F(|\vec p_f-\vec p_i +\vec\kappa +(\vec k'-\vec \kappa)|)
\approx F(|\vec p_f-\vec p_i +\vec\kappa |)+(\vec k'-\vec \kappa)\cdot
\vec\nabla_{q_2}F(q_2),
\end{eqnarray} 
where $q_2=|\vec p_f-\vec p_i +\vec\kappa |$. 
Using this in Eq.~(\ref{e:is}) leads to the appearance of 
$\vec k'$ in the numerator 
of the integral, which
can again be replaced by a gradient on $\vec r$. The result is
\begin{equation}
I_s(\vec k,\vec r) \approx \left(1-F(q_2)\right)J_{on}(\vec k,\vec r)
+ \delta I_s(\vec k,\vec r)
\label{e:isa}
\end{equation}
with 
\begin{equation}
\delta I_s(\vec k,\vec r) \equiv 
-e^{-i\vec k\cdot\vec r}
\vec\nabla_{q_2}F(q_2)\cdot \vec V_{on}(\vec k,\vec r).
\label{e:isb}
\end{equation}

The leading long distance behavior of the integral of Eq.~(\ref{e:vondef}) 
required to evaluate $\vec V_{on}$ is that of 
$e^{i\omega r}$. But 
 $(-i {\vec\nabla_r}-\vec\kappa)
e^{i\omega r} =0$. Furthermore, $\vec\nabla_{q_2}F(q_2)$
is of order $(n / q_2) F(q_2)$, with $n\approx 4$.
Thus, 
the $\delta I_s(\vec k,\vec r)$ screening correction term of Eq.~(\ref{e:isa})
provides a correction which has
an extra factor of $F(q_2)/ q_2 r$ compared to the leading term.
However, F($q_2$) is very small, 1\% at most, 
 for the kinematics of this experiment. 
For typical kinematics $q_2\approx 100$ keV
so $(n / q_2 r) F(q_2)\approx (4/125) (1/100)\approx 3\times 10^{-4}$.
The correction to the separated atom approximation due to the 
screening term is completely negligible here. This is shown in Fig.~6.

\subsection{Coulomb photon propagation}
If we 
keep the $\vec k'$ in the Coulomb photon propagator $1/|\vec q\; '|^2$, we 
need to evaluate
the Coulomb integral $I_{coul}(\vec k,\vec r)$:
\begin{equation}
I_{coul}(\vec k,\vec r) = \int 
\; {d^3k'\over (2\pi)^3} 
\; e^{-i(\vec k - \vec k')\cdot\vec r}
\; \biggl\{
{1\over |\vec p_f -\vec p_i+\vec k'|^2}\; {F(|\vec k -\vec k'|)\over 
\omega^2-\vec k'^{2} + i\epsilon}\biggr\} \;.
\label{e:ic}
\end{equation}
In this case, there are two sets of poles. One is the usual one
at $|\vec k'|=\omega$, but there is also a set of poles off the real axis 
(in the complex $|\vec k'|$ plane) corresponding to the zeros of 
$|\vec p_f -\vec p_i+\vec k'|^2$. 
It is desirable to handle these pole terms separately, 
so we use the identity 
\begin{eqnarray}
{1\over A}\cdot{1\over B}\equiv \left[{1\over 
A}+{1\over B}\right]{1\over A+B} 
\end{eqnarray}
in the form
\begin{eqnarray}
{1\over |\vec p_f -\vec p_i+\vec k'|^2} \cdot 
{1\over \omega^2-\vec k'^{2} + i\epsilon}=
\left[{1\over \omega^2-\vec k'^{2} + i\epsilon} + 
{1\over |\vec p_f -\vec p_i+\vec k'|^2}
\right]\nonumber\\
 {1\over \omega^2 + ( \vec p_i-\vec p_f)^2+2\vec k'\cdot(\vec p_f - \vec p_i)}.
\label{e:cdecomp}
\end{eqnarray}

The first term has the pole at $|\vec k'|= \omega$ 
which is responsible for the long 
distance photon propagation. The second term has the above 
mentioned poles of  $|\vec k'|$ off the real axis.
The vanishing of the denominator 
$\omega^2 + ( \vec p_i-\vec p_f)^2+2\vec k'\cdot(\vec p_f - \vec p_i)$ 
occurs only 
when the two terms in the bracket cancel and causes no 
mathematical difficulty.
The first term of Eq.~(\ref{e:cdecomp}) can be approximated by 
using $\vec k'=\vec\kappa + (\vec k'-\vec\kappa)$ and 
expanding so that
\begin{eqnarray}
{1\over \omega^2-\vec k'^{2} + i\epsilon} \cdot {1\over 
 \omega^2 + ( \vec p_i-\vec p_f)^2+2\vec k'\cdot(\vec p_f - \vec p_i)}
\approx {1\over \omega^2-\vec k'^{2} + i\epsilon} 
\ {1\over (\vec p_f - \vec p_i+\vec\kappa)^2} 
\nonumber\\
\left[1-{2 (\vec k'-\vec\kappa)\cdot
 (\vec p_f - \vec p_i)\over(\vec p_f - \vec p_i+\kappa)^2}\right].
\label{e:ft}
\end{eqnarray}

The first part of Eq.~(\ref{e:ft}) corresponds to the separated atom
approximation. The next term involves $(\vec k'-\vec\kappa)$ which 
yields the integral $V_{on}(\vec k, \vec r \;)$ of Eq.~(\ref{e:vondef}).
The specific correction  to $  J_{on}$ is denoted as
$\delta J_{on}^{coul}$ which is 
obtained by keeping the second term of Eq.~(\ref{e:ft}) in the $I_c$
integral given by Eq.~(\ref{e:is}) so that
\begin{equation}
\delta J_{on}^{coul}(\vec k, \vec r) \equiv (-2)
\int {d^3k'\over (2\pi)^3} 
e^{-i(\vec k - \vec k')\cdot\vec r}
\biggl\{ 
{ ({\vec k \; '} - {\vec \kappa}) \cdot ( \vec p_f - \vec p_i) 
\over 
{\omega^2 - \vec k'^{2} + i\epsilon} }
\biggr\}.
\label{e:deltaic}
\end{equation}
The $ {\vec k \; '} - \vec \kappa$ term of Eq.~(\ref{e:deltaic}) again leads to 
an extra factor of $1/q_2 r$, as compared to the leading term.
The expected small size of this correction is confirmed by 
numerical evaluation. Indeed the second term of
Eq.~(\ref{e:ft}) is negligible except for 
the 
angles for which the leading 
term vanishes. See Fig.~7, which shows 
the relative sizes of 
$J_{on}(\vec k, \vec r)/(\vec p_f - \vec p_i+\vec\kappa)^2$
and the correction to it due to 
$\delta J_{on}^{coul}(\vec k, \vec r)/(\vec p_f - \vec p_i+\vec\kappa)^2$.

What about the second term of Eq.~(\ref{e:cdecomp})? This is given by
$$ \left[ {1\over |\vec p_f -\vec p_i+\vec k'|^2} \right]
{1\over \omega^2 + ( \vec p_i-\vec p_f)^2+2\vec k'\cdot(\vec p_f - \vec
 p_i)}.$$
 It is necessary to treat the $\vec k'\cdot\vec k'$
terms correctly, but the term $ {\vec k \; '} \cdot (\vec p_f - \vec p_i)$
may be evaluated by using
$\vec k'= \vec \kappa +(\vec k'-\vec \kappa)$ and treating 
the difference term as a perturbation in $\vec k'-\vec \kappa$. Thus 
\begin{eqnarray}
{1\over |\vec p_f -\vec p_i+\vec k'|^2}
\approx 
{1\over (\vec p_f - \vec p_i)^2 
+2\vec\kappa\cdot(\vec p_f - \vec p_i)+\vec k'\cdot
\vec k'}
-{2(\vec k'-\vec \kappa)
\cdot (\vec p_f - \vec p_i)
\over ( (\vec p_f - \vec p_i)^2
+2\vec\kappa\cdot(\vec p_f - \vec p_i)+\vec k'\cdot
\vec k')^2}
\label{e:a1}
\end{eqnarray}
 If we treat this 
as a  function of $|\vec k'|$, the poles in 
the exact 
expression and in its approximation given by the first term, appear at
the same positions. The second term can be thought of as correcting the 
value of the residue at the pole. Furthermore, it vanishes for well-separated 
atoms. Thus neglecting the second term
is a good approximation.
We use  similar logic to write
\begin{eqnarray}
{1\over\omega^2 + 
( \vec p_i-\vec p_f)^2+2\vec k'\cdot(\vec p_f - \vec p_i)}\approx
{1\over\omega^2 + ( \vec p_i-\vec p_f)^2+2\vec \kappa\cdot(\vec p_f - \vec
p_i)}
\label{e:a2}
\end{eqnarray}

We use the first terms of Eqs. (\ref{e:a1}) and  (\ref{e:a2})
to  estimate the second term of 
Eq.~(\ref{e:cdecomp}). We immediately expect that 
this second term is completely negligible because it has the form of the 
Fourier transform of 
$${1\over \vec k'\cdot \vec k' + P^2},$$ 
where 
$P^2=(\vec p_f - \vec p_i)^2 +2\vec \kappa\cdot (\vec p_f - \vec p_i).$
This  Fourier transform falls off very rapidly with $r$, {\it{i.e.,}} as
${\exp({-P r}) / r}$. 
For typical values of
P of about $100~ keV \approx 50 {\rm{\AA}}^{-1}$, we will have $Pr
\approx 25$,  where a is the separation ($ circa~2.56~{\rm{\AA}} $). 
The exponential damping factor destroys this second term.
This intuitive conclusion is also confirmed 
by numerical evaluation, but the strong exponential damping inherent in
this term caused the effects of this correction to be too small to be plotted.

\subsection{Electron propagation}

The full expression for the Feynman graphs in Fig.~2 allows a new type 
of term, one in which the electron propagates over the distance $r$.
The mathematical origins of this effect are in the 
poles of the electron propagator shown in 
 Eq.~(\ref{e:twelve}) which arise via 
the appearance of the four-momentum of the virtual
$k^{\;' \; 2}\ne 0$ in those denominators. There are 
two terms in that equation, one arising from the uncrossed graph (Fig.~2a)
and the other from the crossed graph (Fig.~2b).
We shall study these terms in sequence, using our standard technique
of writing $\vec k'=\vec\kappa +(\vec k'-\vec\kappa)$ and treating the second 
term as an expansion parameter---whenever possible without 
destroying the correct analytic structure.

\subsubsection{Uncrossed term}

Suppose we keep the full uncrossed term. This means that we must evaluate 
the integral
$J_1(\vec k, \vec r \;)$:
\begin{equation}
J_1(\vec k, \vec r \;) \equiv \int {d^3k'\over (2\pi)^3}\;
{e^{-i(\vec k-\vec k') \cdot\vec r}\over 2p_f\cdot k'+k'^{2}+i\epsilon}\;
{ 1 \over k'^{2}+i\epsilon}\;
{F(|\vec k -\vec k'|)\over \omega^2-\vec k'^{2} + i\epsilon}\; .
\label{e:eighteen}
\end{equation}
The product of propagators can be written 
\begin{eqnarray}
{1\over 2p_f\cdot k'+k'^{2}} \cdot {1\over k'^{2}+i\epsilon}= \nonumber\\
\left[{1\over k'^{2}+i\epsilon}-{1\over 2p_f\cdot k'+k'^{2}}\right]
{1\over 2p_f\cdot k'} \; .
\label{e:decomp2}
\end{eqnarray}
The ${1 / (k'^{\; 2}+i\epsilon)}$ is the photon propagator and we 
denote its contribution as
the photon propagation term; similarly,  the second part
is the electron propagation term. The last factor 
${1 /2p_f\cdot k'}$ vanishes only when the term in the 
bracket vanishes and so causes no mathematical difficulty.
We may then study two separate integrals
\begin{eqnarray}
J_1(\vec k, \vec r \;)= K_1(\vec k, \vec r \;)+ K_2(\vec k, \vec r \;),
\end{eqnarray}
where 
\begin{eqnarray}
K_1(\vec k, \vec r \;) \equiv \int 
{d^3k'\over (2\pi)^3}
\;{e^{-i(\vec k-\vec k') \cdot\vec r}\over 2p_f\cdot k'}
\;{1 \over k'^{2}+i\epsilon} 
\; {F(|\vec k -\vec k'|)\over \omega^2-\vec k'^{2} + i\epsilon}
\label{e:k1}
\end{eqnarray}
and
\begin{eqnarray}
K_2(\vec k, \vec r \;) \equiv -\int  
{d^3k'\over (2\pi)^3}
\; {e^{-i(\vec k-\vec k') \cdot\vec r}\over 2p_f\cdot k'+k'^{2}+i\epsilon}
\; {1 \over 2p_f\cdot k'} 
\; {F(|\vec k -\vec k'|)\over \omega^2-\vec k'^{2} + i\epsilon} \; .
\label{e:k2}
\end{eqnarray}
We work first with $K_1$. The manipulations are simplified 
by using 
\begin{eqnarray}
{1\over 2p_f\cdot k'}={1\over 2p_f\cdot \kappa}+
{1\over 2p_f\cdot \kappa} 2p_f\cdot (\kappa-k')
{1\over 2p_f\cdot k'}
\label{e:nineteen}
\end{eqnarray}
where 
the four vector $\kappa \equiv (\omega,\vec \kappa)$.
Using this relation in Eq.~(\ref{e:k1})
 enables us to derive a differential equation for $K_1$:
\begin{eqnarray}
K_1(\vec k, \vec r \;) = {J_{on}(\vec k, \vec r \;)\over 2p_f\cdot \kappa} 
-{1\over 2p_f \cdot \kappa} 2\vec p_f\cdot \left[\vec \kappa-
{\vec \nabla\over i}-\vec k\right] K_1(\vec k, \vec r \;).
\label{e:k1eq}
\end{eqnarray}
We see that the first term is the separated atom approximation for this
particular term.
 The quantity
$\left[\vec \kappa +
{ i \vec \nabla_r}-\vec k\right]$ vanishes when acting on
$e^{-i\vec k\cdot\vec r} e^{i\omega r} / r $, so that the second term is
 a correction. The effect of this term can be estimated 
by  replacing $K_1$ on the right hand side by 
${J_{on} / (2p_f \cdot \kappa})$. 
Thus 
\begin{eqnarray}
(2 p_f \cdot \kappa) \; K_1(\vec k, \vec r \;) \approx 
J_{on}(\vec k, \vec r \;) 
+ \delta J_{on}^{uncr} (\vec k, \vec r \;), 
\end{eqnarray}
with 
\begin{eqnarray}
\delta J_{on}^{uncr} (\vec k, \vec r \;) \equiv - 2 \vec p_f \cdot 
\left[ \vec \kappa
+ i {\vec \nabla_r}-\vec k 
\right] 
J_{on}(\vec k, \vec r \;) \; .
\label{e:deltajuncrossed}
\end{eqnarray}

A brief calculation shows that once
again the correction is proportional to 
the vector integral $\vec {V}_{on}(\vec k, \vec r)$ of Eq.~(\ref{e:vondef}),
 and is down by 
about ${1 / \omega r}$ compared to the first term. 
Explicit numerical evaluation confirms this estimate, the correction term 
$J_{on}^{uncr}$ is indeed negligible, as shown in Fig.~8.
For general purposes, it is useful to note that Eq.~(\ref{e:k1eq})
has the formal solution
\begin{eqnarray}
K_1(\vec k, \vec r \;)={1
\over 2p_f\cdot\kappa +2\vec p_f\cdot (\vec \kappa-\vec k
+i{\vec \nabla_r})} J_{on}(\vec k, \vec r \;).
\label{e:k1solv}
\end{eqnarray}
This formal solution
gives us a controlled way to study some of
the corrections to the separated atom approximation.

The $K_2$ term given by Eq.~(\ref{e:k2}) represents new physics occuring in
this  two-atom process. To see this, recall that  $2p_f\cdot k'
+k^{\;'2}+i\epsilon = (p_f+k')^2-M^2+i\epsilon =(E_f  +\omega)^2-m^2- (\vec
p_f+\vec k')\cdot(\vec p_f+\vec k')+i\epsilon$. We write this in terms of a 
four vector W=$(E_f+\omega,\vec p_f+\vec k')$ as $2p_f\cdot k'
+k^{\;'2}+i\epsilon =W^2-m^2+i\epsilon$.
The zero in this term is a pole in the integrand representing the long
distance propagation of the electron of four-momentum $W_{on}=(E_f+\omega,
|\vec p_f+\vec k|\hat r)$. We use the same pole dominance 
idea 
that we used for the photon propagation terms. In this case, the four-vector W 
appears instead of the four vector $\kappa$. Then we handle the 
term $2p_f\cdot k'=2p_f\cdot W -2m^2$ by using 
\begin{eqnarray}
{1\over p_f\cdot W-m^2}={1\over p_f\cdot W_{on}-m^2} +
{1\over p_f\cdot W_{on}-m^2}p_f\cdot\left(W_{on}-W\right)
{1\over p_f\cdot W-m^2}.
\end{eqnarray}

This allows us to derive the analogous differential equation for $K_2$:
\begin{eqnarray}
K_2(\vec k, \vec r \;)={K_{on}(\vec k, \vec r \;)\over 2 p_f\cdot W_{on}-m^2}
-{1\over 2 p_f\cdot W_{on}-m^2} 2\vec p_f\cdot \left(\vec W_{on}
-({\vec{\nabla}\over i}+\vec k+\vec p_f)\right)K_2(\vec k, \vec r \;).
\label{e:k2sol}
\end{eqnarray}
where
\begin{eqnarray}
K_{on}(\vec k, \vec r \;) \equiv -\int {d^3k'\over (2\pi)^3}
\; e^{-i(\vec k-\vec k') \cdot r}
\;{1 \over 2p_f\cdot k'+k^{\;'2}}
\; {F(|\vec k -\vec k'|)\over \omega^2-\vec k'^{2} + i\epsilon} \; .
\label{e:k2on}
\end{eqnarray}
Eq.~(\ref{e:k2sol}) is equivalent to the full expression for $K_2$ and 
also shows
how we can make a first approximation for $K_2$ by substituting
${K_{on}(\vec k, \vec r \;) / (2 p_f\cdot W_{on}-m^2})$ for $K_2$ on the right 
hand side.  The technique is the same as in previous sections.
We immediately see that the second term vanishes in the separated 
atom approximation.

We may evaluate $K_{on}$ in the separated atom approximation, 
because this is essentially the same integral as $J_{on}$. The separated 
atom approximation
worked except when $q_1$ was small. Here the quantity 
$| \vec k-|\vec P_f+\vec k| \; \hat r  |$ plays the same role as $q_1$.
The result is 
\begin{eqnarray}
K_{on}(\vec k, \vec r \;)
 = e^{-i\vec k\cdot\vec r}e^{-i\vec p_f\cdot\vec r} & & \left
(-{1\over 4\pi r}\right)e^{i\sqrt{2E_f\omega+\omega^2+p^2_f}r} 
F\left(|\vec k+\vec p_f -\sqrt{E_f\omega+\omega^2+p^2_f}
\hat r|\right) .
\label{e:twenty-five}
\end{eqnarray}
Since $E_f = \sqrt{m^2+p^2_f}$ is very large, F is evaluated
with a large argument, and this kills the $K_2$ term.

The size of the quantity $K_2$ is controlled by $K_{on}$ and by the 
denominator
\begin{eqnarray}
D\equiv 2 p_f\cdot W_{on}-m^2 =
E_f (E_f+\omega)- (\vec p_f\cdot\hat r) \; |\vec p_f+\vec k|
\end{eqnarray}
Numerical evaluation shows that $K_2\ll K_1$. 
If $\vec p_f = 0$, $K_{on}$ is small because the atomic form factor is 
evaluated at a large argument. When $|\vec p_f|$ takes on a typical 
experimental value, the energy denominator is very large.
The net result is that $K_2$ is ignorable.

However when  $|\vec p_f|$ is very much larger than the electron 
mass, D approaches 0 and $K_2$ can become 
large. The bremsstrahlung from a collection of atoms would then have a large 
contribution from the electron propagation term.
This small value of D is a necessary condition for the occurance of the
Landau-Pomeranchuk-Migdal (LPM) effect 
\cite{lpm} in which the long time scale of electron 
propagation allows a coherent effect which reduces the radiation.
However, we are concerned with the low energy limit in which 
the electron momentum is much less than its mass. So, for us, the electron 
propagation term is negligible.

\subsubsection{Crossed term}

Suppose we keep the full crossed term. This means that we must evaluate 
the integral $J_2(\vec k, \vec r \;)$:
\begin{equation}
J_2(\vec k, \vec r \;) \equiv \int {d^3k'\over (2\pi)^3}
\; {e^{-i(\vec k-\vec k') \cdot\vec r}\over 2p_i\cdot k'-k'^{2}+i\epsilon}
\;{1 \over k'^{2}+i\epsilon}
\; {F(|\vec k -\vec k'|)\over \omega^2-\vec k'^{2} + i\epsilon} \;.
\label{e:eighteenc}
\end{equation}
The product of propagators can be written 
\begin{eqnarray}
{1\over 2p_i\cdot k'-k'^{2}} \cdot {1\over k'^{2}+i\epsilon} = 
\left[{1\over k'^{2}+i\epsilon}+{1\over 2p_i\cdot k'-k'^{2}}\right]
{1\over 2p_i\cdot k'}\; .
\label{e:decomp3}
\end{eqnarray}

Again we denote the first term as 
the photon propagation term and the second term
as the electron propagation term. The last factor 
${1 /( 2p_i\cdot k')}$ vanishes only when 
the term in the bracket vanishes, so this
zero residue pole makes no contribution to the integral 
$J_2(\vec k, \vec r \;)$.
We may then study two separate integrals
\begin{eqnarray}
J_2(\vec k, \vec r \;)= K_3(\vec k, \vec r \;)+ K_4(\vec k, \vec r \;),
\end{eqnarray}
where 
\begin{eqnarray}
K_3(\vec k, \vec r \;) \equiv \int {d^3k'\over (2\pi)^3}
\; {e^{-i(\vec k-\vec k') \cdot\vec r}\over 2p_i\cdot k'}
\; {1 \over k'^{2}+i\epsilon} 
\; {F(|\vec k -\vec k'|)\over \omega^2-\vec k'^{2} + i\epsilon}
\label{e:k3}
\end{eqnarray}
and
\begin{eqnarray}
K_4(\vec k, \vec r \;) \equiv \int {d^3k'\over (2\pi)^3}
\; {e^{-i(\vec k-\vec k') \cdot\vec r}\over 2p_i\cdot k'-k'^{2}+i\epsilon}
\; {1 \over 2p_i\cdot k'} 
\; {F(|\vec k -\vec k'|)\over \omega^2-\vec k'^{2} + i\epsilon} \; .
\label{e:k4}
\end{eqnarray}

It is clear that we can handle $K_3$ using the same techniques 
that we used for $K_1$.
We 
derive the following differential equation for $K_3$:
\begin{eqnarray}
K_3(\vec k, \vec r \;) = {J_{on}(\vec k, \vec r \;)\over 2p_i\cdot \kappa} 
-{1\over 2p_i\cdot\kappa} 2\vec p_i\cdot \left[\vec \kappa-
{\vec \nabla_r\over i}-\vec k\right] K_3(\vec k, \vec r \;).
\label{e:k3eq}
\end{eqnarray}

Once again the first term is the separated atom approximation for this
particular term. This dominates $K_3$.
For general purposes, it is useful to note that Eq.~(\ref{e:k3eq})
has the formal solution
\begin{eqnarray}
K_3(\vec k, \vec r \;)
={1\over 2p_i\cdot\kappa +2\vec p_i\cdot (\vec \kappa-\vec k
+i{\vec \nabla_r})} J_{on}(\vec k, \vec r \;).
\label{e:k3solv}
\end{eqnarray}
Again, this formal solution
gives us a controlled way to study the 
corrections to the separated atom approximation.
We use 
\begin{eqnarray}
(2 p_i \cdot \kappa) K_3(\vec k, \vec r) = J_{on}(\vec k, \vec r)
+ \delta J_{on}^{cr} (\vec k, \vec r) 
\end{eqnarray}
with 
\begin{eqnarray}
\delta J_{on}^{cr}(\vec k, \vec r) \equiv
-2 \vec p_i \cdot \left[ \vec \kappa + i {\vec \nabla_r} - \vec k \right]
 J_{on} (\vec k, \vec r) 
\label{e:deltajcrossed}
\end{eqnarray}
We find the 
cross term correction effects due to
$\delta J_{on}^{cr}$
are very small, as illustrated in Fig.~9.

The term $K_4$ of Eq.~(\ref{e:k4}) represents new physics occuring in this 
two-atom process. We use the same techniques we used for $K_2$.
There is a pole in the integrand representing the long
distance propagation of the electron of four-momentum $X_{on}=(E_i-\omega,
|\vec p_i-\vec k|\hat r)$. 
This allows us to derive the differential equation
\begin{eqnarray}
K_4(\vec k, \vec r \;)={L_{on}(\vec k, \vec r \;)\over 2 p_i\cdot X_{on}+m^2}
-{1\over 2 p_i\cdot X_{on}-m^2} 2\vec p_i\cdot \left(\vec X_{on}
-({\vec{\nabla_r}\over i}+\vec k+\vec p_i)\right)K_4(\vec k, \vec r \;).
\label{e:k4sol}
\end{eqnarray}
where 
\begin{eqnarray}
L_{on}(\vec k, \vec r \;) \equiv -\int {d^3k'\over (2\pi)^3}
\; e^{-i(\vec k-\vec k') \cdot r}
\; {1 \over 2p_i\cdot k'-{k'}^2}
\; {F(|\vec k -\vec k'|)\over \omega^2-\vec k'^{2} + i\epsilon} \; .
\label{e:lon}
\end{eqnarray}
Note the appearance of the $+m^2$ term in the denominator of the first term 
of $K_4$. This renders the energy denominator very large, it never vanishes 
even for infinitely large $E_i$. Careful 
numerical evaluation leads to negligibly small 
results for  $K_4$. These results are too small to be plotted.

\subsection{Summary of Bremsstrahlung Cross Section with Thompson Scattering}

The basic expression for the bremsstrahlung holography 
cross section where the object atom scattering can be described by the Thompson
amplitude  is given by Eqs.~(\ref{e:nine})
and (\ref{e:ten}). The full on- and off-shell integral $I(\vec k, \vec r \;)$ of
Eq.~(\ref{e:iint}) is very well approximated by the on-shell
integral $I_{on}(\vec k, \vec r \;)$
of  Eq.~(\ref{e:fourteen}).

Furthermore, $\kappa_\mu\kappa^\mu=0$ so that we may perform the sum
over the polarization vectors $\hat \epsilon$ (see Ref.~\cite{bd}, Eq.~(7.61)) 
with the results
\begin{equation}
\sum_\epsilon \hat\epsilon \cdot \vec {V}(k) \hat\epsilon\cdot\vec{ V}(k)=
{2 p_f\cdot  p_i\over p_f\cdot k p_i\cdot k}
-{m^2\over (p_f\cdot k)^2}-{m^2\over (p_i\cdot k)^2}
\end{equation}
and 
\begin{equation}
\sum_\epsilon \hat\epsilon \cdot \vec {V}(k) \hat\epsilon\cdot\vec{ V}(\kappa)
={ p_f\cdot  p_i\over p_f\cdot k \;p_i\cdot \kappa}
+{ p_f\cdot  p_i\over p_f\cdot \kappa \;p_i\cdot k}
-{m^2\over p_f\cdot k p_f\cdot \kappa}
-{m^2\over p_i\cdot k p_i \cdot \kappa} \; .
\end{equation}
The net result is that 
\begin{eqnarray}
\sum_\epsilon \overline{|{\cal M}|^2}
& = &
      \left({Ze^3\over |\vec q\;|^2}\left(1-F(|\vec p_f - \vec p_i +
      \vec k|)\right)\right)^2 
      \left[{2 p_f\cdot  p_i\over p_f\cdot k p_i\cdot k}  
      -{m^2\over (p_f\cdot k)^2}-{m^2\over (p_i\cdot k)^2}\right]
\nonumber \\
& + & 
      {Z^3e^8\over |\vec q\;|^2}\; (1-F(|\vec q \; |))
      {(1-F(|\vec p_f-\vec p_i+\vec\kappa|))
      \over|\vec p_f-\vec p_i+ \vec\kappa|^2}
      {2\over m}\; Re \;J_{on}(\vec k, \vec r \;) 
\nonumber \\
& \times &
\left[{ p_f\cdot  p_i\over p_f\cdot k p_i\cdot \kappa}
+{ p_f\cdot  p_i\over p_f\cdot \kappa p_i\cdot k}
-{m^2\over p_f\cdot k p_f\cdot \kappa}
-{m^2\over p_i\cdot k p_i \cdot \kappa}\right] \; .
\label{e:ttten}
\end{eqnarray}
The factors in the square brackets account for the peaking of the 
bremsstrahlung radiation intensity which occurs 
in the direction of the initial electron velocity.
This feature represents the influence of the vector nature of the photon and 
is significantly different than the simple classical result  
from scalar electrodynamics.

The polarization dependent cross section is obtained simply from 
Eq.~(\ref{e:ten}).
by replacing the 
term $I(\vec k,\vec r)$  with $I_{on}$ of 
Eq.~(\ref{e:fourteen}). That is,  we find
\begin{eqnarray}
\overline{|{\cal M}(\epsilon)|^2} 
& \equiv & 
\left({Ze^3\over |\vec q\;|^2}\left(1-F(|\vec p_f -
\vec p_i + \vec k|)\right)\right)^2 
\; [\hat\epsilon \cdot \vec {V}(k)
\;  \hat\epsilon\cdot\vec{ V}(k)] 
{Z^3e^8\over |\vec q\;|^2}\;{2(1-F(|\vec q \; |))\over m}\; 
Re \; I_{on}(\vec k,\vec r) \; .
\label{e:tenon}
\end{eqnarray}

\section{Bremsstrahlung x-ray Holography (BXH) including resonant scattering}

We now return to the case where the scattering of 
the bremsstrahlung photon by the object atoms includes
the resonance correction terms 
given by Eqs.~(\ref{e:rdef}),~(\ref{e:rsimp}), and ~(\ref{e:sdef}).
We simply use those equations in the expression for the amplitude
given by Eq.~(\ref{e:four}). 
The total Thompson plus resonant hologram amplitude ${\cal M}_{T+R}$
is the sum of the Born approximation reference term ${\cal R}$
and the Thompson plus resonant object amplitude $\cal O$.
Calculating this amplitude and squaring
leads to the resonance correction 
$\overline{|{\delta{\cal M}}_R|^2}$ to the non-resonant Thompson 
squared matrix element
$\overline {|{\cal M}|^2}$ with
\begin{eqnarray}
\overline{|{\delta{\cal M}}_R|^2} \equiv 
& + & {Z^2e^6(1-F(|\vec q \; |))\over |\vec q\;|^2}\;S(\omega)\; 2\; Re\;
\hat\epsilon \cdot \vec {V}(k) \int {d^3k'\over (2
\pi)^3}
\; \hat\epsilon\cdot\vec{ V}_1(k')
\; e^{-i(\vec k-\vec k')\cdot\vec r} \nonumber \\
& \times &  
\biggl\{{\left(1-F(|\vec q\; '|)\right)\over |\vec q\; '|^2}\;{F\left(|\vec k
-\vec k'|\right)\over\omega^2-\vec k'^{2} + i\epsilon}\biggr\}\;,
\label{e:rint}
\end{eqnarray}
where $S(\omega)$ is given by Eq.~(\ref{e:sdef}).
If the resonant scattering and the Thompson scattering terms are of comparable
strength, the full 
square of the resonant and non-resonant hologram amplitude is 
the sum of the terms given by  Eqs.~(\ref{e:ttten}) and (\ref{e:rint}):
\begin{eqnarray}
\overline {|{\cal M}_{T+R}|^2} = 
\overline {|{\cal M}|^2} + \overline{|{\delta{\cal M}}_R|^2} \; .
\end{eqnarray}

The integral over $d^3k'$
is the same as that evaluated in the previous sections. The essential 
result is that the integral can be evaluated by
removing  the expression 
$\vec{ V}_1(k')\left(1-F(|\vec q\; '|)\right)/ |\vec q\; '|^2$ and  evaluating 
it for $k'=\kappa$. Recall that $\kappa=(\omega,\omega\hat r)$.
In that case,
\begin{equation}
\overline {|{\delta{\cal M}}_R|^2}=
 +  {Z^2e^6\;\left(1-F(|\vec q \; |)\right)
\over |\vec q\;|^2}\;S(\omega)\; 2\; Re\; J_{on}(\vec k, \vec r \;)
\; \hat\epsilon \cdot \vec {V}(k) 
\; \hat\epsilon\cdot\vec{V}(\kappa)
{(1-F(|\vec p_f-\vec p_i+\vec\kappa|))\over|\vec p_f-\vec p_i+ \vec\kappa|^2}.
\label{e:rint1}
\end{equation}
As explained above, the diagrams with propagating electrons 
are also potentially troublesome---at high energies these effects produce
the interesting LPM effects.
Here 
the off-shell electron effects are governed by known atomic
wave functions which go into computing the function $S(\omega)$.

The sum over the polarization vectors of the photon leads to the
expression
\begin{eqnarray}
\sum_\epsilon \overline{|{\delta{\cal M}}_R|^2}=
& + & {Z^2e^6\;\left(1-F(|\vec q \; |)\right)
\over |\vec q\;|^2}\;S(\omega)\; 2\; Re\; J_{on}(\vec k, \vec r \;)
\; {(1-F(|\vec p_f-\vec p_i+\vec\kappa|))\over|\vec p_f-\vec p_i+ \vec\kappa|^2}
\nonumber \\
& \times & 
\left[{ p_f\cdot  p_i\over p_f\cdot k \ p_i\cdot \kappa}
+{ p_f\cdot  p_i\over p_f\cdot \kappa \ p_i\cdot k}
-{m^2\over p_f\cdot k \ p_f\cdot \kappa}
-{m^2\over p_i\cdot k \ p_i \cdot \kappa}\right]. 
\label{e:rint11}
\end{eqnarray}

\section{X-ray Fluorescence Holography (XFH)}

We now briefly consider the physics of x-ray fluorescence holography (XFH).
In this case, an atom in the sample is excited from its ground state into
an excited state 
by an incoming photon or an incoming electron.  After the ionization,
the excited atom decays and we must consider the interference effects
for the outgoing fluorescence photon via the direct path to the detector 
and via the single-scattering paths to the detector.  

The Feynman diagrams for XFH are shown in Fig.~10.
Suppose the incident photon (Fig.~10a)---or the incident
electron (Fig.~10b)---interacts with 
an atom, knocking an s-shell electron into a
continuum  wave function c. A p-shell electron can 
spontaneously decay to the s-state, emitting  
a fluorescence photon with the characteristic energies
of the atom $\omega=E_p-E_s$, 
and with the reference amplitude 
${\cal R}_{psf \; (c \rightarrow i)} (\vec k,\hat\epsilon)$
for photostimulated fluorescence, 
or with the reference amplitude
${\cal R}_{esf \; (c \rightarrow i)} (\vec k,\hat\epsilon)$
for electron stimulated fluorescence.

First, consider the ionization process.  For photoionization, the 
incoming photon is real or on-shell.  For electron induced ionization,
the incoming electron is on-shell and the ionization occurs via
the virtual photon exchange, but the potential off-shell photon effects 
are exactly the same as the  bremsstrahlung case we have  already analyzed 
in detail. For the relatively low energies used in the current XFH experiments, 
these effects are completely negligible.

Second, consider the intermediate states.
What are the possible off-shell electron
effects?  The electron promoted into the continuum is detectable and is
therefore  on-shell.
 The virtuality of the continuum electron 
enters only if there is another  final state
interaction; such 
effects are of higher order in $\alpha$ and are neglected here.
Thus the only possible off-shell electron effects come from the virtual 
intermediate state electron in the s-state. But this is governed by
well known atomic wavefunctions.
 
For the characteristic radiation used in XFH,
the long wavelength approximation holds and the radiation is dominated by
the electric dipole process.  This predominately dipole character,
combined with the fact that 
the initial and final  electron is on-shell, indicates immediately that  
the separated atom approximation will be extremely close to the
exact quantum electrodynamic solution. 
Thus the holographic reference amplitude takes the simple 
form
\begin{equation}
{\cal R}_{psf \; (c \rightarrow i)} (\vec k,\hat\epsilon)
  = \omega \; T_{psf} (\vec k,\hat\epsilon)
\end{equation}
for photoionization, and 
\begin{equation}
{\cal R}_{esf \; (c \rightarrow i)} (\vec k,\hat\epsilon)
  = \omega \; T_{esf} (\vec k,\hat\epsilon)
\end{equation}
for electron ionization,  
where the $T_{psf}(\vec k,\hat\epsilon)$ and $T_{esf}(\vec k,\hat\epsilon)$ 
factors account for the
remainder of the atomic matrix element.

The holographic object amplitude 
contribution to the total amplitude occurs 
because the photon is scattered coherently by the object atoms.  
When the Thompson rescattering effects dominate, 
the on-shell approximations for the FXH holographic interference terms
are given by the expressions:
\begin{equation}
{\cal M}_{psf}^{on}(\vec k,\hat\epsilon) =
\omega \; T_{psf} (\vec k,\hat\epsilon) 
\left(1 -{e\over m}J_{on}(\vec k, \vec r \;)\right)
\label{e:psfxh}
\end{equation}
for photon stimulated XFH, and
\begin{equation}
{\cal M}_{esf}^{on}(\vec k,\hat\epsilon) =
\omega \; T_{esf} (\vec k,\hat\epsilon) 
\left(1 -{e\over m}J_{on}(\vec k, \vec r \;)\right)
\label{e:esfxh}
\end{equation}
for electron stimulated XFH.
Here the integral $J_{on}(\vec k, \vec r)$ is given by 
Eqns.~(\ref{e:fourteena}) and (\ref{e:exact}). 

The cross section is obtained by squaring the amplitude and 
summing over the polarization vectors of the photon. Thus the 
cross section for the intensity of the XFH hologram
in the on-shell approximation  
is given generically by 
\begin{equation}
{d\sigma\over d\Omega} = \omega^2 \; |T(\vec k,\hat\epsilon)|^2 \;
\left(1-2{e\over m}Re \; J_{on}(\vec k, \vec r \;)\right). 
\end{equation}
And again, when the high electron density regions of the atoms are 
sufficiently well separated, we will recover the usual classical 
holographic form in the far field limit via Eq.~(\ref{e:seventeen}).

\section{Multiple Energy X-Ray Holography (MEXH)}

Finally, we consider very briefly the physics of multiple energy x-ray
holography  (MEXH).  In this case, a real photon is sent into the sample from
outside and we  must consider the interference between the direct path to the
detector atom and the  single scattering paths via the object atoms to the
detector atom.  The Feynman diagrams for this interference are shown in
Fig.~11. Note that the Feynman diagrams  for MEXH are not just the time
reversed Feynman diagrams for photon induced XFH, and that there are three
interfering terms in MEXH.

Within classical electrodynamics, MEXH has been related to XFH by the
reciprocity theorem, which can be paraphrased roughly as follows:  Put the
source outside the sample and the detector inside the sample, turn on the
source, and measure the electric field at the detector;   if the positions of
the source and the detector are interchanged, the electric field measured at
the detector will be the same. This result comes from the time reversal
invariance of Maxwell's equations.  

How does this very reasonable classical result emerge from the quantum 
electrodynamic treatment?  It clearly is not just simple time reversal
invariance, since there are three interfering diagrams in MEXH and only two
interfering diagrams in XFH.  
There are three diagrams
in MEXH because the incoming photons interfere to produce the atomic
excitations that lead to the fluorescence, and the outgoing fluorescence
photons interfere just as they do in XFH.  However since MEXH 
must average over
many outgoing directions to increase the signal level, 
the interference effects in the outgoing 
photons will be washed out \cite{photonexafs},
 and we need only consider the first two diagrams
in Fig.~11. Then our question becomes how are these two diagrams related to
the analogous diagrams for XFH shown in Fig.~10. They still are not just the
simple 
time reversed diagrams: in MEXH the Thompson process (or, in general, the 
Compton process) occurs in the incoming state of the photon that will produce 
the photoionization,  whereas in XFH the Thompson process is in the outgoing 
photon state of energy $(E_p-E_s)$ that will be detected.
Since the incoming photon energy used in MEXH is not equal to 
$(E_p-E_s)$, the two processes are not related by time reversal invariance.

We discuss this further by displaying the relevant equation.
In MEXH, the holographic object amplitude 
contribution to the total amplitude occurs 
because the initial photon in the incoming beam with momentum $k_b$ 
and polarization $\hat \epsilon_b$ is
scattered elastically by the object atoms prior to absorption by 
the detector atom.  
If the total matrix element to produce the outgoing angle averaged MEXH 
fluorescence decay is  denoted 
${\cal R}_{me \; (c \rightarrow i)} (\vec k_b,\hat\epsilon_b)$
and if
the Thompson scattering effects dominate, then 
the on-shell approximation for the holographic interference term
is given by the expression
\begin{equation}
{\cal M}_{me}^{on} (\vec k_b,\hat\epsilon_b) =
\left(1 - {e\over m}J_{on}(\vec k_b,\vec r)\right)\; 
\omega \; T_{me}(\vec k_b,\hat\epsilon_b), 
\label{e:mexh}
\end{equation}
where 
$\omega T_{me}(\vec k_b,\hat\epsilon_b)$ represents 
the photon absorption process,
and once again the integral $J_{on}(\vec k_b, \vec r)$ is given by 
Eqns.~(\ref{e:fourteena}) and (\ref{e:exact}). 
The amplitude for MEXH given by Eq.~(\ref{e:mexh}) and the amplitude for
FXH given by Eqs.~(\ref{e:psfxh}) and (\ref{e:esfxh}) are not complex
conjugates of one another. 
However, the physics of these two amplitudes is closely
related.

\section{Summary and Discussion}

We have shown that, if the energy of the bremsstrahlung photons is measured,
the bremsstrahlung radiation produced inside a
crystal will produce a holographic interference pattern in the 
far field outside the crystal.  To use this new form of generalized holography
\cite{porter},
we must know the reference and object amplitudes.
These amplitudes were calculated using 
quantum electrodynamics, and compared with 
the corresponding predictions of classical scalar electrodynamics.  
The essential results for 
bremsstrahlung holography are displayed in 
Eqs.~(\ref{e:ttten}), (\ref{e:tenon}), and (\ref{e:rint11}).

The total amplitude
$\cal M$ is the sum of the reference amplitude $\cal R$ and the
object amplitude $\cal O$.  
To obtain very accurate results, the full expression for 
the quantity $J_{on}(\vec k, \vec r \;)$ given by Eq.~(\ref{e:exact}) must be
used. 
Its simpler asymptotic form given by Eq.~(\ref{e:seventeen}) is not
accurate for the case in which the
separation $\vec r$ between the source atom and the object atom
is parallel to the direction $\hat k$ of the detected photon, as shown in 
Fig.~5.

The key feature in obtaining 
Eqs. (\ref{e:ttten}) and  (\ref{e:rint11})
is that the 
photons that propagate from the source atom to the object atoms are essentially 
on-shell---the square of their four momenta is very close to zero. 
Sect. V is devoted to the
detailed arguments for our on-shell separated atom approximation.
We show explicitly that all of the known
short-ranged  off-shell virtual effects are negligibly small for the
low ($40-60$ keV) electron energies used in the current experiments.
The underlying reason for this is that the atoms in solids
are  too far apart for the off-shell photons or electrons, 
produced via bremsstrahlung or via fluorescence, to
propagate from one atom to another.

It is interesting to compare the present case in which photons
propagate between atoms with two examples from nuclear physics:
(1) hadronic scattering
from nuclei,
and (2) pion production in nucleon-nucleon or
nucleon-nucleus collisions.
Beg's theorem \cite{Beg} applies to hadron-nuclear scattering and 
states
that, if the target nucleons are separated by distances greater
than the range of the hadron-nucleon interaction,
then the hadron-nucleus scattering amplitude can be expressed in terms
of on-shell hadron nucleon scattering amplitudes. This is 
called the separated scatterer
approximation\cite{ssa1,ssa2,ssa3,ssa4}. In this language,
our results can be stated as the confirmation that the 
analogous separated atom approximation is valid.

In our first example, we want to consider the scattering of hadrons
from the nucleons inside the nucleus.  In our condensed matter
physics example, we were able to consider the scattering of photons 
from essentially stationary atoms and slowly moving electrons.  
However, in nuclei,
 the nucleons move extremely rapidly so that the separations
between the nucleons fluctuate 
and, in addition, the nucleons can overlap. The average separation
distance between nucleons is about $1.8$ fm,
 which is about twice as big as
the typical range of hadron-nucleon interactions ({\it{circa}} 1 fm). However,
in the case of very high energy hadronic beams, 
we can be reasonably sure that the nucleons will not move during the
passage of
the hadron through the target, and that the use of on-shell hadron
nucleon amplitudes based on the 
average separation  distance is valid

In our second example, the pion ($\pi$) production reactions
pp $\to d\pi$  or p + A $\to (A+1)\pi^+$,
are the strong 
interaction analogs of bremsstrahlung holography. In this case,
Fig. 2 also applies, but the wiggly lines represent pions and the
solid vertical lines represent nucleons or nuclei. These
processes involve high momentum transfer (for pions produced with
low energy) and small inter-nucleon separations.  Therefore
off-shell pions can and do propagate between nucleons. As a
result, no cross sections can be computed to 
better than about a factor of two  \cite{measday}.
In contrast, in the present paper,
we can calculate cross sections reliably at the 10$^{-4}$ level or better
because the atoms
and the electrons in the atoms move relatively slowly.

The separability argument works extremely well for the three x-ray 
holographies (FXH, MEXH and BXH) currently under experimental development. 
In a  sense, it is
quantum electrodynamics that requires the separated atom approximation to work
so well since it supplies the forces responsible for the 
relatively slow motion of the
atoms and of the electrons in the atoms,
and it also supplies the interactions between the 
atoms and the incoming and outgoing photons and electrons.

Thus it is quantum electrodynamics that gives us our main results for
bremsstrahlung holography summarized in 
Eqs.~(\ref{e:nine}) through (\ref{e:twelve}), Eq.~(\ref{e:ttten}),
and Eq.~(\ref{e:rint11}). Quantum
electrodynamics shows 
that bremsstrahlung holography should work---the remaining
problems are experimental.

\section{Acknowledgements}

We gratefully acknowledge the partial support of this work by 
the US Department of Energy
under Grants DE-FG03-97ER41014 and DE-FG06-88ER40427,
by the Japanese New Energy and Industrial Technology Development Organization,
and by the University of Washington Research Royality Fund under
Grant 65-9976. 
One of us (GAM) thanks the national Institute for Nuclear Theory for
partial support during the completion of this work. 

We thank Prof. John Rehr for supplying us with the isolated atom
and embedded atom electron densities for crystalline copper shown 
in Fig.~4.

\newpage

\noindent{\bf{Figure Captions}}
\bigskip

1. The classical scalar wave description of internal source holography.
The source atom produces the spherical reference wave $R$ which 
propagates directly to the detector, and which is
scattered by the object atom to produce the spherical object wave $O$.  
The interference between $R$ and $O$ at the far field detector produces 
the internal source hologram.
\medskip

2. The uncrossed (a) and crossed (b) Feynman diagrams for the reference
amplitude  $\cal R$ in bremsstrahlung x-ray holography (BXH). The x represents
the target atom  that produces the  bremsstrahlung photon.  In the uncrossed
(crossed) diagram,  the outgoing photon is created after (before) the virtual
photon is destroyed. The solid line represents the incident electron and the
wiggly lines the photons;   this standard convention is used in all the Feynman
diagrams in this paper.
\medskip
 
3. The four Feynman diagrams for the BXH object amplitude ${\cal O}$.  The
object amplitude has four terms due to the crossed and uncrossed source terms,
and the  crossed and uncrossed Compton scattering terms.  The sum of the two
$\cal R$ diagrams in Fig.~2 interfere with the sum of the four ${\cal O}$
diagrams in this fi
gure to produce the bremsstrahlung hologram.
\medskip

4. The electron density $\rho (s)$ for isolated copper atoms (dashed line)
and for copper atoms in crystalline copper (solid line) calculated using the
FEFF computer code \cite{Rehr}.  Here $s$ is the distance from the center of
the atom.
\medskip

5. Comparison of the real part of the
on-shell separated atom approximation $J_{on}$
(solid line) given by Eq.~(\ref{e:fourteena})
with the real part of the
classical spherical wave holography function (dashed line)
given by Eq.~(\ref{e:seventeen}).  
The results in Figs. 5--9 are shown for crystalline
copper with representative experimental kinematics: the incident electron
energy is 60 keV and the outgoing photon energy is 20 keV.   
Here $\cos (\theta) = \hat k \cdot \hat r $.
\medskip

6. The small effects of photon virtuality on the screening correction
given by the ratio  of the second to the first term in Eq.~(\ref{e:isa}). 
The ratio $Re\; \delta I_s /  Re \; [(1-F(q_2))\; J_{on}]$ is plotted to
illustrate the size of these corrections for two typical experimental
values of the momentum transfer, namely 12.2 keV and 93.5 keV.  
Here the momentum transfer $\vec\Delta\equiv \vec p_i-\vec p_f$ and the
angle $\theta'$ is specified by $cos(\theta')=\hat\Delta\cdot\hat r$.
\medskip

7. The small effects of virtual photon propagation on the Coulomb correction.
The real part of the on-shell Coulomb correction 
$\delta J_{on}^{coul}/(\vec p_f - \vec p_i+\vec\kappa)^2$ 
(dashed line) given by Eq.~(\ref{e:deltaic}) is compared with the
real part of the full on-shell separated atom approximation 
$J_{on}/(\vec p_f - \vec p_i+\vec\kappa)^2$ 
(solid line) given by Eq.~(\ref{e:fourteena}). 
Here $\Delta = 12~{\rm{keV}}$,
$\hat p_i\cdot\hat r=0.5$, and $\hat p_f\cdot\hat p_1$=0.5.
\medskip

8. The small effects of virtual electron propagation in the uncrossed graph in
Fig.~2.  The real part of the uncrossed correction $\delta J_{on}^{uncr}$ 
(dashed line) given by Eq.~(\ref{e:deltajuncrossed}) is compared with the 
real part of the on-shell separated atom approximation $J_{on}$ (solid line) 
given by Eq.~(\ref{e:fourteena}). Here $\hat k \cdot \hat p_f = 0.5$
and $\hat p_i \cdot \hat r=0.5$.
\medskip

9. The small effects of virtual electron propagation in the crossed graph in
Fig.~2.  The real part of the crossed correction $\delta J_{on}^{cr}$ 
(dashed line) given by Eq.~(\ref{e:deltajcrossed}) is compared with the 
real part of the on-shell separated atom approximation $J_{on}$ (solid line) 
given by Eq.~(\ref{e:fourteena}). Here $\hat p_i\cdot\hat r=0.5$.
\medskip

10. The Feynman diagrams for x-ray fluorescence holography (XFH).
The s-state 
core hole can be made by photoionization (a), or by electron induced
ionization (b).  The black dot represents the photon--object atom  scattering
amplitude, 
c the continuum electron, p the  p-state electron,  and s the s-state
electron.  Note that the black dot is shorthand for two diagrams---namely, the 
crossed and uncrossed Compton diagrams shown in Fig.~3.
\medskip

11. The Feynman diagrams for multiple energy x-ray holography (MEXH).
The notation is the same as Fig.~10.


\begin{thebibliography} {99}
\bibitem{so_many} There are so many papers on electron holography that it
is impossible for us to provide all the references.  Since our paper is
on x-ray holography, we have tried only to provide an entry into the 
literature based on recent papers.  We certainly
do not mean to imply that the recent papers we have quoted for this 
purpose properly give credit for the extensive work in electron holography.
\bibitem{dynamical} When the crystal is nearly perfect, we must use the dynamic
theory of x-ray diffraction instead of the kinematic theory.
In the dynamical scattering limit, the incoming particles coherently
multiply scatter from many atoms in the crystal. The interference between these
many multiple-scattering events produces the Bragg-Darwin peaks.
\bibitem{photonexafs} However, even the complete $4\pi$ outgoing angle averaged
MEXH will still contain a subtle modulation which is the 
photon equivalent of XAFS, but these photon-XAFS-like oscillations in the angle 
averaged MEXH fluorescence will be very small and can be neglected in 
experimental MEXH. This 
photon-XAFS provides only spherically averaged distances 
like electron-XAFS.
\end{thebibliography}
\end{document}